\documentclass[a4paper,11pt]{article}

\usepackage{graphicx,array}
\usepackage{color}
\usepackage{latexsym}
\usepackage{amsthm}
\usepackage{amsmath}
\usepackage{amssymb}
\usepackage{empheq}

\usepackage{dsfont}
\usepackage{epsfig}
\usepackage{slashed}
\usepackage{bbold}
\usepackage{psfrag}
\usepackage[svgnames]{xcolor}
\PassOptionsToPackage{caption=false}{subfig}
\usepackage{subcaption}
\usepackage{xfrac}
\usepackage{multirow}
\usepackage{booktabs}
\usepackage{cite}
\usepackage[normalem]{ulem}
\usepackage{jheppub} % for details on the use of the package, please see the JHEP-author-manual

\newcommand{\be}{\begin{equation}}
\newcommand{\ee}{\end{equation}}
\newcommand{\bea}{\begin{eqnarray}}
\newcommand{\eea}{\end{eqnarray}}
\newcommand{\bei}{\begin{itemize}}
\newcommand{\eei}{\end{itemize}}

\newcommand{\nn}{\nonumber}

\title{The see-saw portal at future Higgs factories:\\
the role of dimension six operators}

\author[a,b]{Daniele Barducci}
\author[c]{Enrico Bertuzzo}

\affiliation[a]{Universit\`a degli Studi di Roma la Sapienza, Piazzale Aldo Moro 5, 00185, Roma, Italy}
\affiliation[b]{INFN Section of Roma 1, Piazzale Aldo Moro 5, 00185, Roma, Italy}
\affiliation[c]{Instituto de Fisica, Universidade de Sao Paulo, C.P. 66.318, 05315-970 Sao Paulo, Brazil}

\emailAdd{daniele.barducci@roma1.infn.it}
\emailAdd{bertuzzo@if.usp.br}

\abstract{We study an extension of the Standard Model with electroweak scale right-handed singlet fermions $N$ that induces neutrino masses, plus a generic new physics sector at a higher scale $\Lambda$. The latter is parametrized in terms of
effective operators in the language of the $\nu$SMEFT.
We study its phenomenology 
considering operators
up to $d=6$, where additional production and decay modes for $N$ are present
in addition to those 
arising from the mixing with the active neutrinos. 
We focus on the production with four-Fermi operators and we identify the most relevant additional decay modes to be $N\to \nu \gamma$ and $N\to 3f$. We assess the sensitivity of future Higgs factories on the $\nu$SMEFT in regions of the parameter space where the new states decay promptly, displaced or are stable on detector lengths. We show that new physics scale up to $5-60\;$TeV can be explored, depending on the collider considered.

}

\begin{document} 
Last update \today
\maketitle

\section{Introduction}

Neutrino masses and mixing can be explained by adding to the Standard Model (SM)  a new Weyl fermion $N$, total singlet under the SM gauge group, which acts as the right-handed (RH) counterpart of the left-handed (LH) SM neutrino. The lightness of the neutrino masses can be explained by the see-saw mechanism~\cite{Minkowski:1977sc,Mohapatra:1979ia,Yanagida:1979as,Gell-Mann:1979vob}
\be\label{eq:seesawscaling}
m_\nu \sim  \frac{y^2 v^2}{M_N} \ ,
\ee
where $v$ is the Higgs vacuum expectation value (VEV), $y$ the strength of the Dirac type interaction and $M_N$ the Majorana mass of the RH neutrino. While there is no indications on the energy scale at which this mechanism takes place, 
there is nowadays a strong interest in models where RH neutrinos have a mass at the EW scale. From one side they are in fact an interesting alternative, in that they can generate the observed matter-antimatter  asymmetry via neutrino oscillations~\cite{Akhmedov:1998qx,Asaka:2005pn}, while they can be searched for at colliders and at beam-dump experiments~\cite{Keung:1983uu,Ferrari:2000sp,Graesser:2007pc,delAguila:2008cj,BhupalDev:2012zg,Helo:2013esa,Blondel:2014bra,Abada:2014cca,Cui:2014twa,Antusch:2015mia,Gago:2015vma,Antusch:2016vyf,Caputo:2016ojx,Das:2017zjc,Caputo:2017pit,Abada:2018sfh,Hernandez:2018cgc,Jones-Perez:2019plk,Beltran:2021hpq}. Moreover, if lighter than ${\cal O}(100)\;$MeV, they can 
be relevant for the solution of longstanding anomalies reported in the neutral-current~\cite{LHCb:2014vgu,LHCb:2017avl,LHCb:2021trn} and charged-current~\cite{BaBar:2012obs,BaBar:2013mob,Belle:2015qfa,Belle:2016ure,Belle:2016dyj,LHCb:2015gmp,LHCb:2017smo,LHCb:2017rln,Belle:2019gij} semileptonic decay of $B$ mesons~\cite{Asadi:2018wea,Greljo:2018ogz,Azatov:2018kzb}. Their phenomenology is driven by the mixing $\theta$ with the active neutrinos
\be\label{eq:theta}
\theta  \sim \frac{y\, v}{M_N} \sim \left(\frac{m_\nu}{M_N} \right)^{1/2}\ ,
\ee 
which drives their production rates and and their decay width and hence their lifetime. 
The naive see-saw scaling of Eq.~\eqref{eq:theta} can be modified if multiple RH neutrinos are present with specific Yukawa and Majorana mass textures that ensure an approximate lepton number symmetry~\cite{Kersten:2007vk,Gavela:2009cd}. Consequently scenarios with a much larger mixing angles can be realized, thus altering the RH neutrinos phenomenology. It's also interesting to speculate on the possibile presence of additional  NP states at a scale $\Lambda \gg v, M_N$, whose effects 
can be parametrized in the language of effective field theories in the so called $\nu$SMEFT, where a tower of higher-dimensional operators $\frac{{\cal O}^d}{\Lambda^{4-d}}$ built out with the SM fields and the RH neutrinos is added to the renormalizable lagrangian.
At the lowest possible dimension, $d=5$, there are two genuine, {\emph{i.e.}} that contains at least a RH neutrino field,
 $\nu$SMEFT operators: one that triggers the decay of the SM Higgs boson into a pair of RH neutrinos and a dipole operator with the hypercharge gauge boson~\cite{Graesser:2007yj,Aparici:2009fh}. Already at $d=6$ many more operators are present~\cite{Graesser:2007pc,delAguila:2008ir,Liao:2016qyd} with interesting phenomenological consequences, since they also can induce new production and decay channels.

Many of these operators have been subject of theoretical studies, especially for what concerns their phenomenology at the Large Hadron Collider (LHC), see {\emph{e.g.}}
\cite{Graesser:2007yj,Graesser:2007pc,Aparici:2009fh,Duarte:2016caz,Caputo:2017pit,Bischer:2019ttk,Alcaide:2019pnf,Butterworth:2019iff,Biekotter:2020tbd,deVries:2020qns,Cottin:2021lzz}. However, RH neutrinos with a mass at the EW scale are one of the primary goals of future lepton colliders, since the generally small production cross section proper of EW singlet states can be overcome, thanks to the clean detector environments and the typically lower SM backgrounds with respect to  hadronic machines. For the post LHC era many future lepton colliders have been proposed. These includes $e^+e^-$ facilities, both circular ones as the Future Circular Collider~\cite{Gomez-Ceballos:2013zzn,Abada:2019zxq,Abada:2019lih,Blondel:2019yqr} (FCC-ee) and the Compact electron-positron collider~\cite{CEPCStudyGroup:2018rmc,CEPCStudyGroup:2018ghi} (CEPC), and linear ones as the International Linear Collider~\cite{Behnke:2013xla,Baer:2013cma,Bambade:2019fyw} (ILC) and the Compact Linear Collider~\cite{deBlas:2018mhx,Roloff:2018dqu} (CLIC).
Finally, a great attention has recently arose for multi TeV $\mu^+\mu^-$ colliders~\cite{Delahaye:2019omf}, which could provide a great handle to test higher-dimensional operators whose effect grows with energy.

In a recent paper~\cite{Barducci:2020icf}, we have investigated the prospects of these machines, commonly denoted as {\emph{Higgs factories}}, in testing the two genuine  $d=5$ operators of the $\nu$SMEFT through Higgs and $Z$ boson physics and  focusing on RH neutrinos with masses in the $[1\;{\rm GeV}-m_{h,Z}]\;$ range. There we have shown that future lepton colliders can test exotic branching ratios (BRs) for the Higgs and $Z$ boson down to $\sim10^{-3}$ and $10^{-9}$ respectively, greatly surpassing the reach of future indirect measurements of the Higgs and $Z$ boson width. In this paper we extend our previous work by studying the phenomenology of the $\nu$SMEFT operators that arise at $d=6$. Since these are typically
generated by different ultraviolet (UV) completions than the $d=5$ ones, the bounds 
on the cut off-scale $\Lambda$ derived in~\cite{Barducci:2020icf}
do not necessarily direct apply\footnote{For related works on the phenomenology of the $\nu$SMEFT at future lepton colliders see {\emph{e.g.}}~\cite{Duarte:2018kiv,Zapata:2022qwo}.
}. 

We focus  on EW scale RH neutrinos with masses in the $[1\;{\rm GeV}-m_W]\;$ range and study the additional production and decay channels induced by the $d=6$ operators. We distinguish two main decay channels: a two body decay into a SM neutrino and a photon, $N\to \nu\gamma$, and a three body decay into a SM leptons and a fermion pair which can proceed either as $N\to \nu f \tilde f$ or $N\to \ell f \tilde f$, where $\ell=e,\mu,\tau$.  In the three body decay cases the final state fermions could involve either a pair of quarks or leptons. For what concerns the production, we identify the most  relevant channels as single-production and pair-production of RH neutrinos induced by four-Fermi $d=6$ operators, since they induce amplitudes that grow with the energy of the process. 

The paper is organized as follows. In Sec.~\ref{sec:framework} we set our notation and review the $\nu$SMEFT framework, while in Sec.~\ref{sec:colliders} we present the properties of the future colliders which are under analysis. Then in Sec.~\ref{sec:decay} we study the main decay channels induced by the $d=6$ operators and present the expressions for the various partial widths. We then show under which conditions these additional decay modes can dominate with respect to the one already present at renormalizable level and induced by the active-sterile mixing. We further quantify the lifetime of the RH neutrinos once these operators are switched on. In Sec.~\ref{sec:prod} we discuss the additional production modes relevant for studies at future Higgs factories. We present our results in Sec.~\ref{sec:prompt}, Sec.~\ref{sec:displaced} and Sec.~\ref{sec:stable} for prompt, displaced and detector stable RH neutrinos. We finally conclude in Sec.~\ref{sec:conclusions}. We 
then report in App.~\ref{sec:amplitudes} the expressions for the spin averaged matrix elements squared relevant for the $N$ three-body decay via an off-shell SM boson induced by $d>4$ operators.

\section{Theoretical framework}\label{sec:framework}

The $\nu$SMEFT is described by the following Lagrangian
\be\label{eq:lag_nusmeft}
{\cal L} = {\cal L}_{\rm SM}  + \bar N \slashed \partial N - \bar L_L Y_\nu \tilde H N - \frac{1}{2} M_{N} \bar N^c N + \sum_{n>4} \frac{{\cal O}^n}{\Lambda^{n-4}} + h.c. \ .
\ee
Here $N$ is a vector describing ${\cal N}$ flavors of RH neutrino fields, singlet under the SM gauge group, and $N^c = C \bar N^T$, with $C= i \gamma^2 \gamma^0$. Furthermore $L_L$ is the SM lepton doublet, $Y_\nu$ is the $3\times {\cal N}$ Yukawa matrix of the neutrino sector with $\tilde H = i \sigma^2 H^*$, $M_{N}$ is a ${\cal N}\times {\cal N}$ Majorana mass matrix for the RH neutrinos  and ${\cal O}^n$ the Lorentz and gauge invariant operators with dimension $n$ built out from the SM and the RH neutrino fields, with $\Lambda$ indicating the EFT cut-off.
In~\cite{Graesser:2007yj,Graesser:2007pc,delAguila:2008ir,Aparici:2009fh,Liao:2016qyd} the $\nu$SMEFT has been built up to $d=7$ and at $d=5$ only three operators exists. The first is the well-know Weinberg operator~\cite{Weinberg:1979sa}
%\be\label{eq:Weinberg}
%{\cal O}^5_{W} = \alpha_W(\bar L^c \tilde H^*)( \tilde H^\dag L ) \ ,
%\ee
while the two genuine $\nu$SMEFT operators are $ {\cal O}^5_{NH} = \alpha_{NH}(\bar N^c N )(H^\dag H )$ and ${\cal O}^5_{NB} = \alpha_{NB}\bar N^c \sigma^{\mu\nu} N B_{\mu\nu}$, where $B_{\mu\nu}$ is the SM hypercharge field strength tensor and $2\sigma^{\mu\nu}=i [\gamma^\mu,\gamma^\nu]$, which have been recently investigated in~\cite{Barducci:2020icf}. At $d=6$ many more operators are present. They are reported in Tab.~\ref{tab:D6_operators}, where we split them between operators that involve the Higgs boson and four-Fermi operators which do not\footnote{For each operator the corresponding Wilson coefficient, which is again a matrix in flavor space, is implied.}.

\begin{table}[t]
\begin{center}
\begin{tabular}[t]{c|c}
\multicolumn{2}{c}{{\bf Operators involving the Higgs boson}} \\
\hline \hline
Operator  & Definition    \\
\hline
${\cal O}_{LNH}^6$ & $(\bar L \tilde H N_R)(H^\dag H)+h.c.$      \\
${\cal O}_{LNB}^6$ & $(\bar L \sigma^{\mu\nu} N_R) B_{\mu\nu} \tilde H+h.c$   \\
${\cal O}_{LNW}^6$ &  $(\bar L \sigma^{\mu\nu} N_R) \sigma^a W_{\mu\nu}^a \tilde H+h.c$   \\
${\cal O}_{NH}^6$      &    $(\bar N_R \gamma^\mu N_R)(H^\dag i \overleftrightarrow{D}_\mu H)$     \\
${\cal O}_{NeH}^6$      &    $(\bar N_R \gamma^\mu e_R)(\tilde H^\dag i \overleftrightarrow{D}_\mu H)+h.c.$     \\
\end{tabular}
\hfill
\begin{tabular}[t]{c|c}
\multicolumn{2}{c}{{\bf four-Fermi vector operators}} \\
\hline \hline 
Operator  & Definition    \\
\hline
${\cal O}_{Ne}^6$      &    $(\bar N_R \gamma^\mu N_R)(\bar e_R \gamma_\mu e_R)$     \\
${\cal O}_{Nu}^6$      &    $(\bar N_R \gamma^\mu N_R)(\bar u_R \gamma_\mu u_R)$     \\
${\cal O}_{Nd}^6$      &    $(\bar N_R \gamma^\mu N_R)(\bar d_R \gamma_\mu d_R)$     \\
${\cal O}_{Nq}^6$      &    $(\bar N_R \gamma^\mu N_R)(\bar q_L \gamma_\mu q_L)$     \\
${\cal O}_{NL}^6$      &    $(\bar N_R \gamma^\mu N_R)(\bar L_L \gamma_\mu L_L)$     \\
${\cal O}_{NN}^6$      &    $(\bar N_R \gamma^\mu N_R)(\bar N_R \gamma_\mu N_R)$    \\
${\cal O}_{Nedu}^6$      &    $(\bar N_R \gamma^\mu e_R)(\bar d_R \gamma_\mu u_R)+h.c.$     \\
\end{tabular}\\
\vskip 15pt
\begin{tabular}[t]{c|c}
\multicolumn{2}{c}{{\bf four-Fermi scalar operators}} \\
\hline \hline 
Operator  & Definition    \\
\hline
${\cal O}_{4N}^6$      &    $(\bar N^c_R N_R)(\bar N^c_R N_R)+h.c.$     \\
${\cal O}_{NLqu}^6$ &  $(\bar N_R L) (\bar q_L u_R)+h.c$    \\
${\cal O}_{LNqd}^6$ &  $(\bar L N_R) \varepsilon (\bar q_L d_R)+h.c$   \\
${\cal O}_{LdqN}^6$ &  $(\bar L d_R) \varepsilon (\bar q_L N_R)+h.c$   \\
${\cal O}_{LNLe}^6$ &  $(\bar L N_R) \varepsilon (\bar L e_R)+h.c$   
\end{tabular}
\hfill
\begin{tabular}[t]{c|c}
\multicolumn{2}{c}{{\bf Other four-Fermi operators}} \\
\hline \hline 
Operator  & Definition    \\
\hline
${\cal O}_{uddN}^6$      &    $(\bar u^c_R d_R \bar d^c_R) N_R+h.c.$     \\
${\cal O}_{qqdN}^6$      &    $(\bar q^c_L \varepsilon q_L \bar d^c_R) N_R+h.c.$     \\
\end{tabular}
\caption{Genuine $d=6$ operators in the $\nu$SMEFT~\cite{Liao:2016qyd}. For each operator the Wilson coefficient, which is a matrix in flavor space, is implied.}
\label{tab:D6_operators}
\end{center}
\end{table}

\subsection{Neutrino mixing formalism}\label{sec:mixing-formalism}

We summarize here the properties of the neutrino sector of the $\nu$SMEFT, and we refer the reader to~\cite{Barducci:2020icf} for a more detailed discussion.
Under the approximation in which the contribution to the active neutrino masses dominates over the ones induced by the effective operators the active neutrino mass matrix takes the standard form
\be\label{eq:nu_mass_def}
m_\nu \simeq \frac{v^2}{2} Y_\nu \frac{1}{M_N} Y_\nu^T = U^* m_\nu^{d} U^\dag \ ,
\ee
where $U$ is the Pontecorvo-Maki-Nakagawa-Sakata (PMNS) matrix~\cite{Pontecorvo:1957qd,Maki:1962mu} and $m_\nu^{d}$ is the diagonal matrix of neutrino masses. Eq.~\eqref{eq:nu_mass_def} can be solved for the Yukawa matrix of the neutrino sector. In the Casa-Ibarra parametrization~\cite{Casas:2001sr} one obtains
\be
Y_\nu \simeq \frac{\sqrt{2}}{v}U^* \sqrt{m}{\cal R}\sqrt{M_N} \ ,
\ee
where $\sqrt m$ is a $3\times {\cal N}$ matrix containing the physical neutrino masses $m_i$ and ${\cal R}$ is a complex orthogonal ${\cal N}\times {\cal N}$ matrix.
We restrict now our study to the case ${\cal N}=2$
 where for the normal hierarchy (NH) and inverted hierarchy (IH) one has for the matrix $\sqrt m$\footnote{When needed, for our numerical estimate we take $m_{\nu_2}=8.6 \times 10^{-3}\;$eV and $m_{\nu_3}=5.1\times 10^{-2}\;$eV for the NH while we take $m_{\nu_1}=4.9 \times 10^{-2}\;$eV and $m_{\nu_2}=5.0\times 10^{-2}\;$eV for the IH.}
\be
\sqrt{m_{{\rm NH}}} = \begin{pmatrix}
0 & 0 \\
0 & \sqrt{m_2} \\
\sqrt{m_3} & 0
\end{pmatrix}\ , ~~~ 
\sqrt{m_{{\rm IH}}} = \begin{pmatrix}
0 & \sqrt{m_1} \\
 \sqrt{m_2}  & 0\\
0& 0
\end{pmatrix} \ ,
\ee
while we parametrize the orthogonal matrix $\mathcal{R}$ in terms of the complex angle $z = \beta + i \gamma$ as
\be\label{eq:Rmatrix}
\mathcal{R} = 
\begin{pmatrix}
\cos z & \pm \sin z \\
- \sin z & \pm \cos z
\end{pmatrix}\ .
\ee
The active-sterile mixing angle is given by
\be
\theta_{\nu N} \simeq - U^* \sqrt{m} {\cal R} \frac{1}{\sqrt{M_N}} \ .
\ee
It's crucial that the angle $z$ is, in general, a complex parameter. In fact, in the limit in which $z$ is a real number, by taking $U$ and ${\cal R}$ with entries of order unity and by assuming an equal value for the diagonal entries of the Majorana mass term for the two RH neutrino  $m_{N_1}=m_{N_2}=m_N$, one obtains the {\emph{naive see-saw scaling}}\footnote{We have assumed NH and fixed $m_\nu = m_{\nu_3}$. The expression holds also for the IH case modulo order one factors.}
\be\label{eq:Ynu_realz}
Y_\nu \sim \frac{\sqrt{m_N m_\nu}}{v} \sim 4 \times 10^{-8} \left(\frac{m_N}{1\;{\rm GeV}} \right)^{1/2} \ .
\ee
This relation is drastically modified by the imaginary part of $z$, that gives an exponential enhancement. In the limit $\gamma \gg 1$ one has
\be
\mathcal{R} \simeq \frac{e^{\gamma- i \beta}}{2}
\begin{pmatrix}
1 & \pm i \\
- i & \pm 1
\end{pmatrix}\ ,
\ee
and the relation of Eq.~\eqref{eq:Ynu_realz} is modified to
\be
Y_\nu \sim  2 \times 10^{-8}  e^{\gamma -  i \beta}\left(\frac{m_N}{1\;{\rm GeV}} \right)^{1/2} \ .
\ee
Clearly the same enhancement is inherited by the active-sterile mixing, that now reads
\be\label{eq:increase}
 \theta_{\alpha i}  \equiv \left(\theta_{\nu N} \right)_{\alpha i} \sim  7.2 \times 10^{-6} \, e^{\gamma-i \beta} \, \left(\frac{1~\mathrm{GeV}}{m_N} \right)^{1/2}\ .
\ee
We use $\alpha = 1,2,3$ for the active neutrino flavor and $i = 1,2$ for the RH neutrino flavor. This deviation from the naive see-saw scaling has a crucial impact on the RH neutrinos phenomenology, especially for what concerns their decay width and consequently their lifetime, with drastic implications for search strategies at future colliders as recently shown in~\cite{Barducci:2020ncz,Barducci:2020icf}.

\begin{table}[t!]
\begin{center}
\scalebox{0.85}{
\begin{tabular}[t]{ c || c | c }
        \multicolumn{3}{c}{{\bf{Higgs run}}}	        \vspace{0.2cm} \\
 Collider & $\sqrt s\;$ [GeV] & $\int {\cal L}\;$[ab$^{-1}$]   \\ 
 \hline
 \hline
  \multirow{1}{*}{FCC-ee} & 240 & 5  \\
\hline
    \multirow{1}{*}{ILC} 	& 250 & 2 (pol)  \\
\hline
       CLIC-380& 380 &1 (pol)   \\	
    \hline	  
        CEPC& 240 &  5.6  \\		
         \multicolumn{3}{c}{{\bf{$Z$ pole run}}}	        \vspace{0.2cm} \\     
Collider & $\sqrt s\;$ [GeV] & $\int {\cal L}\;$[ab$^{-1}$]  \\ 
 \hline
 \hline
  \multirow{1}{*}{FCC-ee} &  $m_Z$ & 150  \\
\hline
        CEPC & $m_Z$ & 16  	             
\end{tabular}
\hspace{1.2cm}
\begin{tabular}[t]{ c   || c | c   }
        \multicolumn{3}{c}{{\bf{High-energy run}}}	        \vspace{0.2cm} \\
 Collider & $\sqrt s\;$ [TeV] & $\int {\cal L}\;$[ab$^{-1}$]  \\ 
 \hline
\hline
        CLIC & $3$ & 3	  \\
        \hline
               \multirow{1}{*}{$\mu\mu$} & $3$ & 1  	\\
\end{tabular}
}
\end{center}
\caption{Center of mass energies and total integrated luminosities for the various collider options considered in the analysis.}
\label{tab:colliders}
\end{table}

\section{Future Higgs factories}\label{sec:colliders}

In this work we study the phenomenology of the $\nu$SMEFT at future Higgs factories, both at their low energy runs, relevant for physics at the Higgs-strahlung threshold and at the $Z$ pole, 
together with high-energy multi TeV runs, which can greatly enhance the sensitivity on higher-dimensional operators which induce a quadratic grow with the energy, as for the case of four-Fermi operators.
For what concerns the low energy runs, various $e^+ e^-$ prototypes, presently at different stages of their design, have been proposed. These include circular ones, as the Future Circular Collider (FCC-ee)~\cite{Gomez-Ceballos:2013zzn,Abada:2019zxq,Abada:2019lih,Blondel:2019yqr}  and the Circular Electron Positron Collider (CEPC)~\cite{CEPCStudyGroup:2018rmc,CEPCStudyGroup:2018ghi}, and linear ones, as the International Linear Collider (ILC)~\cite{Bambade:2019fyw,Behnke:2013xla,Baer:2013cma} and the Compact Linear Collider (CLIC)~\cite{deBlas:2018mhx,Roloff:2018dqu}. 
Regarding colliders in the multi TeV regime, prototypes include CLIC with a center of mass energy of $3\;$TeV~\cite{deBlas:2018mhx,Roloff:2018dqu} and a  $\mu^+\mu^-$ colliders with various center of mass and luminosity options~\cite{Delahaye:2019omf}. We report in Tab.~\ref{tab:colliders} the main parameters of these colliders prototypes. For concreteness and clarity of presentation, in this work we will present our results only for the case of FCC-ee at $\sqrt s=m_Z$ and $\sqrt s=240$\;GeV, for a $\mu\mu$ collider at $\sqrt s=3\;$TeV  and for CLIC at $3$ TeV, since the ILC and CEPC prototypes will have an overall similar behavior to the considered options.

\section{Decay channels for RH neutrinos}\label{sec:decay}

%%%
%%%

\begin{table}[t!]
\begin{center}
\scalebox{0.95}{
\begin{tabular}[t]{cccc}
Operator & Decay & Mixing & Loop  \\
\hline\hline
Mixing			  &  $N \to 3f$ & \checkmark & $\times$  \\
\hline
\multirow{2}{*}{${\cal O}^5_{NB}$}	& $N_2 \to N_1\gamma $ & $\times$ & \checkmark  
 \\
							& $N\to \nu \gamma$	& \checkmark & \checkmark 	
							 \\
\hline
\multirow{3}{*}{${\cal O}^6_{LNB,W}$}& $N \to \nu\gamma $ & $\times$ & \checkmark   \\
 & $N \to \nu Z^* $ & $\times$ & \checkmark   \\
 & $N \to \ell W^* $ & $\times$ & \checkmark   \\
\end{tabular}
}
\hfill
\scalebox{0.95}{
\begin{tabular}[t]{cccc}
Operator & Decay & Mixing & Loop  \\
\hline\hline
 ${\cal O}^6_{LNH}$ & $N\to \nu H^*$  & $\times$ & $\times$  \\
    ${\cal O}^6_{NH}$ &$N\to \nu Z^*$  & $\checkmark$ & $\times$ \\
   ${\cal O}^6_{NeH}$ &$N\to \ell W^*$  & $\times$ & $\times$ \\
\hline
${\cal O}^6_{4f}$ $-$ {\emph{neutral}} & $N \to 3f$ & $\checkmark$ & $\times$  \\
${\cal O}^6_{4f}$ $-$ {\emph{charged}} & $N \to 3f$ & $\times$ & $\times$  
\end{tabular}
}
\end{center}
\caption{ Decay modes for the RH neutrinos induced by higher-dimensional operators and renormalizable mixing. We highlight whether the corresponding rates are mixing and/or loop suppressed. {\emph{Neutral}} and {\emph{charged}}  indicate the four-Fermi operators with two and one RH neutrino respectively.}
\label{tab:rates}
\end{table}

%%%
%%%

At the renormalizable level, the RH neutrinos only decay thanks to the mixing with their SM active counterpart, a pattern which is not altered by the inclusion of $d=5$ operators except in the case of a 
sufficiently large mass splitting  $m_{N_2}- m_{N_1}$ (with $m_{N_2}>m_{N_1}$) in which the ${\cal O}^5_{NB}$ operator can trigger a non-negligible $N_2 \to N_1 \gamma$ decay rate~\cite{Barducci:2020icf}.
The inclusion of $d=6$ operators can dramatically alter this behavior, leading to new decay patterns, see also~\cite{Duarte:2015iba,Duarte:2016miz}. 
For example, the four-Fermi operators reported in Tab.~\ref{tab:D6_operators} can induce the decay of a RH neutrino into three SM fermions, $N\to 3f$. 
Depending on the operator, the rate for this decay
may or may not be suppressed 
by the active-sterile mixing angle. In particular, it is suppressed in the case of four-Fermi operators which contain two RH neutrino fields, while unsuppressed otherwise.
On the other side, the operators involving the Higgs boson can induce, after EW symmetry breaking, the decay into a final state fermion and a massive SM boson, $B=Z, W^\pm, h$. Given the 
range of
RH neutrino masses on which we are interested in, the SM boson turns out to be off shell and the resulting decays are thus $N\to \nu B^*$ and $N\to \ell B^*$, with the subsequent decay $B^* \to f \tilde f$. Also in this case the rate could be, or not, suppressed by the active-sterile mixing and again the final state is composed by three SM fermions, as for the case of the four-Fermi operators. However  the kinematic and the flavor composition is generally different. Finally, the SM boson could be a massless photon, and the decay is thus simply $N \to \nu \gamma$. We now discuss the various operators and the decay that they mediate in turn, summarizing their main properties in Tab.~\ref{tab:rates}, where we highlight whether the decays that they generate are suppressed by mixing and/or loop effects. We then report in App.~\ref{sec:amplitudes} the spin averaged amplitude squares for the considered three body decays. 

\subsection{Operators that induce $N\to \nu \gamma$}\label{sec:decay_na}

This decay mode is induced at the $d=5$  level by the ${\cal O}^5_{NB }$ operator and at the $d=6$ level by the ${\cal O}^6_{LNB,LNW}$ operators.

\subsubsection*{Decay from ${\cal O}^5_{NB}$}
This operator gives the $N_i\to \nu_\alpha\gamma$ decay only after a mixing insertion\footnote{We assume that the two RH neutrinos are (almost) degenerate and that the $N_2 \to N_1\gamma$ decay rate is negligible.}. The rate for this decay reads~\cite{Aparici:2009fh}
\be
\Gamma(N_i \to \nu_\alpha \gamma) = \frac{\left|\alpha_{NB}\right|^2}{(16\pi^2)^2}\frac{2 c_\omega^2}{\pi} \frac{m_N^3}{\Lambda^2}  \left| \theta_{\alpha i} \right|^2 \ ,
\ee
where $c_\omega$ is the cosine of the Weinberg angle and where we have explicitly introduced a loop suppression factor, since in any weakly coupled UV completion this operator arises at loop level~\cite{Buchmuller:1985jz,Craig:2019wmo}. 

\subsubsection*{Decay from ${\cal O}^6_{LNB,LNW}$}

The operators  ${\cal O}^{6}_{LNB,LNW}$ induce the decay $N_i\to \nu_\alpha \gamma$ with a rate~\cite{Butterworth:2019iff}
\be\label{eq:nugamma}
\Gamma(N_i \to \nu_\alpha \gamma)=
\frac{|(\alpha_{LNB}^6)_{\alpha i} c_\omega + (\alpha_{LNW}^6)_{\alpha i} s_\omega|}{(16\pi^2)^2}^2\frac{v^2}{4\pi} \frac{m_N^3}{\Lambda^4}  \ ,
\ee
where, again, we have explicitly written the loop suppression factor. This decay is not suppressed by the active-sterile mixing.

\subsection{Operators that induce $N\to 3f$}\label{sec:decay_n3f}

This decay mode has contribution from both operators involving the Higgs boson as well as four-Fermi operators. Also the decay induced at the renormalizable level by the active-sterile mixing produces the same final state.

\subsubsection*{Decay from ${\cal O}^6_{LNH}$}

This operators induces the decay $N_i\to \nu_\alpha H^*$ and the final rate is thus suppressed by the Yukawa couplings of the SM fermions to the Higgs boson, but is unsuppressed by the active-sterile mixing. By neglecting phase space effects due to the masses of the final state fermions and working at  ${\cal O}(m_H^{-4})$ the decay rate reads
\be
\Gamma(N_i \to \nu_\alpha f\bar f) =  \frac{3\,N_c}{4096 \pi^3} \left|(\alpha_{LNH}^6)_{\alpha i}\right|^2 \left(\frac{m_f}{v}\right)^2 \left(\frac{v}{\Lambda}\right)^4 \frac{m_N^5}{m_H^4}  ,
\ee 
where the factor $N_c=3$ is present if the final state is a quark-antiquark pair. In our numerical analysis we use the full expression for the decay and sum over the relevant $f\bar f$ final states for any $N$ mass. For the decay into a $b\bar b$ final state, which is the relevant one for $m_N > 10\;$GeV, one has
\be
\Gamma(N \to \nu b\bar b)  \sim   7.5 \times 10^{-12} \left(\frac{1\;{\rm TeV}}{\Lambda}\right)^4  \left(\frac{m_N}{30\;{\rm GeV}}\right)^5\;{\rm GeV} 
\ee
where the Wilson coefficient has been fixed to one.

\subsubsection*{Decay from ${\cal O}^6_{NH}$}

This operator induces the decay $N_i\to \nu_\alpha Z^*$ and the rate is suppressed by a mixing insertion. For a generic $f\bar f$ final state arising from the $Z^*$ decay one has, in the limit $m_f=0$ and at ${\cal O}(m_Z^{-4})$ 
\be
\Gamma(N_i \to \nu_\alpha f\bar f) = \frac{N_c }{3072\pi^3} \left| (\alpha^6_{NH})_{ij} \theta_{j\alpha} \right|^2
\frac{e^4}{ s_w^4 c_w^4} (g_L^2+g_R^2)\left(\frac{v}{\Lambda} \right)^4   \frac{m_N^5}{m_Z^4} \ ,
\ee
where
\be
g_L =t_3 - q s_w^2 \ , \qquad  g_R = -q s_w^2 \ ,
\ee
with $t_3=\pm 1/2 $ and where $q$ is the electric charge of the final state fermion pair. For example for the decay into a final state bottom pair, where $t_3=-1/2$ and $q=-1/3$ one has
\be
\Gamma \sim 2.3 \times 10^{-9} \theta^2 \left(\frac{1\;{\rm TeV}}{\Lambda}\right)^4 \left(\frac{m_N}{30\;{\rm GeV}}\right)^5\;{\rm GeV} \ ,
\ee
where $\theta$ schematically indicates the relevant mixing angle.

\subsubsection*{Decay from ${\cal O}^6_{NeH}$}

This operator induces the mixing unsuppressed decay $N_i\to \ell_\alpha W^*$. Working again in the limit where all the final state fermions are massless, the decay for one of the two charge conjugate modes reads
\be
\Gamma(N_i \to \ell_\alpha f f') = \frac{N_c |(\alpha_{NeH}^6)_{i\alpha}|^2}{6144\pi^3} \left(\frac{g v}{\Lambda} \right)^4  \frac{m_N^5}{m_W^4}  \sim 5.4 \times 10^{-9} \left(\frac{1\;{\rm TeV}}{\Lambda}\right)^4 \left(\frac{m_N}{30\;{\rm GeV}}\right)^5\;{\rm GeV}
\ee
where the estimate is with $N_c=3$ and an ${\cal O}(1)$ Wilson coefficient.

\subsubsection*{Decay from ${\cal O}^6_{LNB,LNW}$}

The combination of these two operators orthogonal to the one that induces $N\to \nu\gamma$ gives again a $N\to \nu Z^*$ decay. In addition, the operator with the $W$ boson produces a $N\to \ell W^*$ decay. Both these rates are not suppressed by the active-sterile mixing. 
In the massless limit, the neutral decay width is
\be
\Gamma(N \to \nu_\alpha f \bar f) = \frac{N_c}{960 \pi^3} \frac{|(\alpha_{LNB}^6)_{\alpha i}\, s_\omega - (\alpha_{LNW}^6)_{\alpha i} \,c_\omega|^2}{(16\pi^2)^2}
\left(\frac{e}{s_\omega c_\omega}\right)^2 (g_L^2+g_R^2)\frac{v^2}{\Lambda^4}\frac{m_N^7}{m_Z^4}
\ee
while for the charged case we obtain
\be
\Gamma(N_i\to \ell_\alpha f \bar f^\prime) =\frac{N_c}{(16\pi^2)^2}\frac{1}{1920\pi^3} \left|(\alpha_{LNW}^6)_{\alpha i} \right|^2 g^2\frac{v^2}{\Lambda^4}\frac{m_N^7}{m_W^4}  \ ,
\ee
where, again, the rate is for one of the two charged conjugated modes.

\subsubsection*{Decay from four-Fermi vector operators: ${\cal O}^6_{Nf}$ and ${\cal O}^6_{Nedu}$}

The first operators are of the form $(\bar N_R \gamma^\mu N_R)(\bar f_{L/R} \gamma_\mu f_{L/R})$. They mediate the decay $N\to \nu f\bar f$ which is suppressed by the active-sterile mixing angle.  For simplicity we assume a diagonal flavor structure for the SM fermion pair final state. In the limit of massless final states the decay rate reads
\be
\Gamma(N_i \to \nu_\alpha f\bar{f}) =    \frac{N_c }{1536 \pi^3} \frac{m_N^5}{\Lambda^4} \left| \theta_{\alpha j} \alpha^6_{ji} \right|^2 \times 2 \ ,
\ee
where $\alpha^6$ is the Wilson coefficient of the four-Fermi operator and the factor $2$ comes from summing over $\nu$ and $\bar \nu$, since also the SM neutrino is Majorana. The charged operator ${\cal O}^6_{Nedu}$ triggers a decay $N \to \ell^- u \bar d + \ell^+ \bar u  d$, which has a rate \\
\be
\Gamma(N_i \to \ell_\alpha u d) =  \frac{N_c\, |(\alpha_{Nedu}^6)_{i\alpha}|^2}{1536 \pi^3} \frac{m_N^5}{\Lambda^4}   \times 2  \ .
\ee

\subsubsection*{Decay from four-Fermi scalar operators}

These operators induce the decay $N_i \to \ell_\alpha f \bar{f}$, where $\ell$ could be a charged or neutral lepton. Each decay proceeds with a rate
\be
\Gamma(N_i \to \ell_\alpha f \bar f) =  \frac{N_c\, |\alpha^6_{i\alpha}|^2}{6144\pi^3}\frac{m_N^5}{\Lambda^4}  \times 2 ,
\ee
where, once more, $\alpha^6$ denotes the generic Wilson coefficient of the four-Fermi operator.

\subsubsection*{Decay via mixing}

Finally, the decay induced via mixing has an approximate rate of~\cite{Atre:2009rg}
\be
\Gamma_{\rm mix}\sim 10^{-2}\left( \frac{m_N}{100\;{\rm GeV}} \right)^5 |\theta_{\alpha i}|^2\;{\rm GeV} \ . 
\ee

\subsection{Which decay dominates?}\label{sec:decay_dominate}

We can now compare the decay rates computed in Secs.~\ref{sec:decay_na} and~\ref{sec:decay_n3f} to see which 
one dominates in the different regions of the $\nu$SMEFT parameter space.
This is essentially determined by three parameters\footnote{While we already stated that we work in the limit where the two RH neutrinos are almost degenerate and the various entries of the active-sterile mixing matrix $\theta$ are determined by the choice of the NH or IH once the RH neutrino mass has been fixed, each higher-dimensional operator can in principle have a different Wilson coefficient. For concreteness, we work under the assumption that they are all equal. We also consider the NH scenario. Results in the IH case are almost identical.}: the mass of the RH neutrino $m_N$, the active-sterile mixing $\theta$ and the EFT cut-off scale $\Lambda$.  We take the latter to be the same for all the considered operators. Clearly, different UV completions will generate at low energy different operators, in general suppressed by different mass scales. We will comment in Sec.~\ref{sec:bound_th} and  Sec.~\ref{sec:bound_exp} on the independent limits on the scale $\Lambda$ that can be set for the most relevant ones. For simplicity however, in performing our main analysis, we will assume that only four fermi operators and the dipole ones triggering the $N\to \nu\gamma$ decay are active, and that they are associated to a unique scale $\Lambda$.

The first question we want to address is in which region of the parameter space 
the decay induced at the renormalizable level by the active-sterile mixing dominates over the decay generated by higher-dimensional operators,
taking into account that current constraints  forces the squared mixing angle to be $\lesssim 10^{-6}$~\cite{Liventsev:2013zz, Aaij:2016xmb, Abreu:1996pa}. In order to do this we need to make some assumptions on the number of four-Fermi operators that are active, since each one can contribute with a multiplicity due to the flavor structure of the operator itself. To be practical, we parametrize this with a coefficient $\xi$ which takes into account how many channels from four-Fermi operators contribute to the decay of a RH neutrino, for example for a decay into a pair of final state quarks $\xi=N_c=3$. Clearly, the most important four-Fermi operators for $N$ decay are the ones that do not pay a mixing suppression, {\emph{i.e.}} 
${\cal O}^6_{Nedu}$ and all the scalar ones.
On the other side, for the operators that contribute to the $3f$ final state via an off-shell $h$, $Z$ and $W$, we can consider all possible decays by summing on their decay modes, since those are fixed by the SM symmetries.
In these calculation we retain the full expressions for the various decay rates.
\begin{figure}[t!]
\begin{center}
\includegraphics[width=0.49\textwidth]{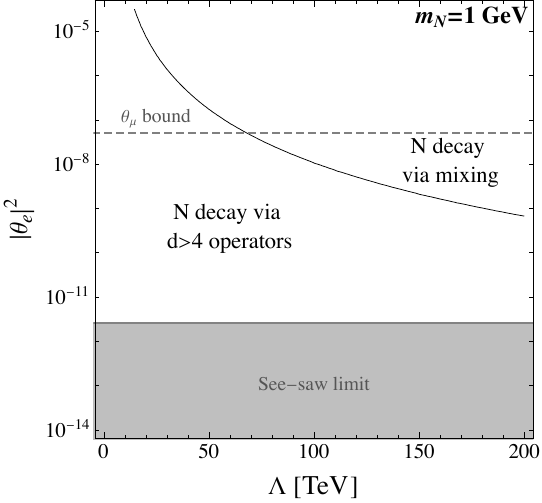}\hfill
\includegraphics[width=0.49\textwidth]{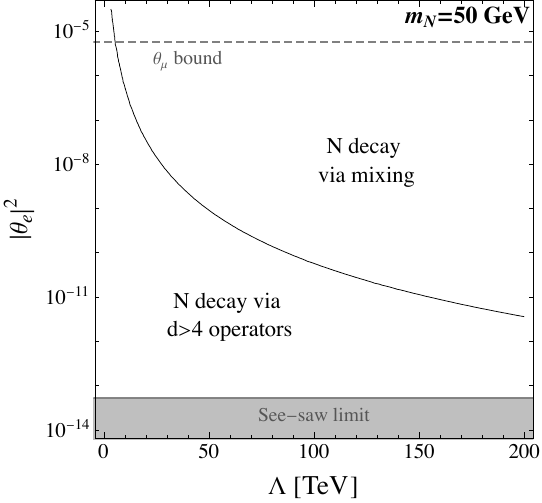}
\caption{\small Dominant decay mode for the RH neutrino in the $\Lambda - |\theta_e|^2$ plane for $m_N=1\;$GeV (left) and $m_N=50\;$GeV (right). The black solid line divides the region where the dominant decay is induced by the mixing or by $d>4$ operators. The gray dashed line indicates the bound on the mixing angle arising from existing experimental searches.
In the gray shaded region the lightness of the neutrino masses cannot be explained by the see-saw mechanism.}
\label{fig:mix_vs_d}
\end{center}
\end{figure}
We then show in Fig.~\ref{fig:mix_vs_d} the region of parameter space where the decay induced by the higher-dimensional operators dominate over the one induced by the mixing. We illustrate them for the degenerate RH neutrino masses $m_N=1\;$GeV (left) and $m_N=50\;$GeV (right)\footnote{The dependence on $\xi$ turns out to be completely negligible for the $N$ mass range of our interest up to $\xi={\cal O}(100)$. }. Above the black solid line the decay pattern is thus the one analyzed in~\cite{Barducci:2020icf}, while effects from higher-dimensional operators become relevant in the lower part of the plot. The dashed gray lines indicate the experimental bound on $\theta_\alpha=\sum_{i=1,2}|\theta_{\alpha i}|^2$ reported in~\cite{Liventsev:2013zz,Aaij:2016xmb,Abreu:1996pa}. We show only the bound on $|\theta_\mu|^2$ which turns out to be the most stringent one. Finally the gray shaded area represents the see-saw limit, below which the lightness of the neutrino masses cannot be explained by the see-saw mechanism. As we see, for small enough $\Lambda$ the dominant decay mode of the RH neutrino can be induced by the higher-dimensional operators of the $\nu$SMEFT while retaining compatibility with existent active-sterile mixing bounds. 
As previously discussed, in this region two decay modes compete: $N\to 3f$, which produces the same final state as the decay via mixing albeit with different kinematics, and $N\to \nu \gamma$. For $|\theta_e|^2\lesssim 10^{-6}$ one has that the ratio $\frac{\Gamma_{N\to \nu\gamma}}{\Gamma_{N\to 3f}}$ is almost independent on $\Lambda$ and that the $\nu\gamma$ decay dominates over the $N\to 3f$ decay for 
 \be\label{eq:na_vs_3f}
m_N \lesssim 15\;{\rm GeV} \ .
\ee
In this region the decay is driven by the $d=6$ operators ${\cal O}^6_{LNB,LNW}$, since the $d=5$ operator ${\cal O}_{NB}$ gives a rate which is mixing suppressed. For larger masses, the operator that dominates the $N \to 3f$ decays is ${\cal O}^6_{NeH}$, which is again not mixing suppressed. 
Given that we are interested in the phenomenology of the $d>4$ operators in the following we will focus in the region where the decay is dominated by higher-dimensional operators and work under the assumption of negligible active-sterile mixing.

\subsection{Bounds from theoretical considerations}\label{sec:bound_th}
The computation of the neutrino properties outlined in Sec.~\ref{sec:framework} rests on the assumption that the $d=4$ masses and Yukawa couplings dominate over the higher dimensional contributions. In order for this to be true, the NP scale will have to satisfy some conditions. Before enumerating them, it is useful to point out that, unlike what happens in the SMEFT, the $\nu$SMEFT is characterized by two expansion parameters: the active-sterile mixing $\theta$ and the cutoff scale $\Lambda^{-1}$. As previously discussed and shown in Fig.~\ref{fig:mix_vs_d}, the phenomenology will strongly depend on the interplay between the two. In order to understand the stability of the $d=4$ parameters against the additional contributions, we will consider only those effects that solely depend on $\Lambda$, neglecting possible effects that are doubly suppressed by some power of $\theta$ and of $1/\Lambda$.

The two $d=5$ operators ${\cal O}^5_{NH}$ and ${\cal O}^5_{NB}$ give a contribution to the RH neutrino mass matrix $M_N$ at tree and 1-loop level respectively. While the contribution from the latter turns out to be irrelevant, the scale suppressing ${\cal O}^5_{NH}$ is bounded by
\be
\Lambda_{NH} \gtrsim  60\;{\rm TeV}\;\frac{1\;{\rm GeV}}{M_N} \ ,
\ee
in order for the $d=4$ contribution to dominate.
Turning to $d=6$ operators, the two main effects come from ${\cal O}^6_{LNH}$, that contributes at tree-level to the neutrino Yukawa coupling, and from ${\cal O}^6_{LNB}/{\cal O}^6_{LNW}$, that contribute at the one-loop level to the active neutrino masses.\footnote{We have explicitly checked that the contribution from the operators ${\cal O}^6_{LNqu}$, ${\cal O}^6_{LNqd}$ and ${\cal O}^6_{LdqN}$ give weaker bounds with respect to the ones shown.} We obtain
\begin{align}\begin{aligned}
& \Lambda_{LNH}  \gtrsim120~{\rm TeV}\left(\frac{10^{-6}}{|\theta|^2} \right)^{\frac{1}{4}}\left(\frac{1\;{\rm GeV}}{M_N} \right)^{\frac{1}{2}} \\
& \Lambda_{LNB,LNW}   \gtrsim 6~{\rm GeV} \left(\frac{10^{-6}}{|\theta|^2} \right)^{\frac{1}{4}}\left(\frac{1\;{\rm GeV}}{M_N} \right)^{\frac{1}{2}}
\end{aligned}\end{align}
where the reference value $|\theta|^2 \sim 10^{-6}$ is the approximate experimental upper bound on the mixing angle for the RH neutrino mass range of our interest. As we can see, the theory bound on the scale of ${\cal O}^6_{LNH}$ is pretty strong, while the one on ${\cal O}^6_{LNB}/{\cal O}^6_{LNW}$ is rather weak, at least for values of the mixing close to the allowed upper bound. In order for the bounds on the scale of these operators to be of the TeV order we would need $|\theta| \sim 10^{-14}$, which is below the see-saw limit, see Fig.~\ref{fig:mix_vs_d}.

\subsection{Bounds from precision measurements}\label{sec:bound_exp}

The operators of Tab.~\ref{tab:D6_operators} involving the Higgs bosons will also trigger additional decay of the SM bosons, which are constrained by precision measurements from LEP and LHC data.
By asking that these additional decay modes do not exceed the absolute uncertainty on the measurement of the $Z$ and $W$ boson width, and that they contribute less than $10\%$ to the SM Higgs boson width, one obtains that the strongest  limit arises from the constraints on $h\to NN$ decay given by ${\cal O}^6_{LNH}$ that reads
\be\label{eq:precision_bound}
\Lambda \gtrsim 5\;{\rm TeV} \ ,
\ee 
a result compatible with the one reported in~\cite{Butterworth:2019iff}. This is due to the small total width of the SM Higgs boson, which compensates for the lower absolute precision on its determination with respect to the $Z$ and $W$ cases. For the latter we obtain a bound of $\Lambda\gtrsim 0.8\;$TeV and 0.6\;TeV, respectively. While for ${\cal O}^6_{LNH}$ the theoretical bound discussed in the previous section is stronger, for the dipole operators the experimental bounds are stronger. 

The interplay between the ${\cal O}^6_{LNB,LNW}$ operator and the active-sterile mixing also generates a magnetic moment $d_\mu \bar \nu \sigma^{\alpha\beta}\nu F_{\alpha\beta}$ for the SM neutrinos which can be estimated as
\be
d_\mu \simeq \frac{1}{16\pi^2} \frac{v}{\Lambda^2}(\alpha_{NB} c_\omega + \alpha_{NW} s_\omega) \theta_{\nu N}^T \ .
\ee 
This is another example of effect which is suppressed by both $\theta$ and powers of $1/\Lambda$. The value of the active-sterile dipole moments constrained by reactor, accelerator and solar neutrino data~\cite{Canas:2015yoa,Miranda:2019wdy} which give
$$
\Lambda \gtrsim 4\times 10^{-2} \left(\frac{|\theta|^2}{10^{-10}} \right)^{1/4} {\rm TeV}.
$$
that is weaker than $\Lambda \gtrsim 1\;$TeV for the allowed mixing angles range.

\subsection{Lifetime of RH neutrinos}\label{sec:lifetime}

After having discussed the main RH neutrinos decay modes, it is important to determine  the lifetime of these state, to assess whether their decay happen with a prompt or displaced behaviour or if instead they are stable on collider lengths. We quantify the three behavior as follows:

\paragraph{Prompt decay}

We consider a RH neutrino to decay promptly if its decay happens within $\sim 0.1\;$cm from the primary vertex. At the renormalizable level, 
prompt RH neutrino decays require a large breaking of the naive see-saw scaling. 
In the notation of Sec.~\ref{sec:mixing-formalism}, this 
is parametrized by a large value of the $\gamma$ parameter, see Eq.~\eqref{eq:increase}. Large mixing angles are however constrained by a variety of experimental searches, and too large values of $\gamma$ are thus ruled out.

\paragraph{Displaced decay}

A particle is considered to decay displaced if it decays away from the primary vertex but within the detector environment. The precise distance for defining a vertex to be displaced clearly depends on the specific detector geometry. Given that our study focuses
on future proposed $e^+e^-$ and $\mu^+\mu^-$ colliders, for which  detailed detector characteristics  have not yet been settled, we consider as displaced particles decaying between  $0.1\,$cm and $1\,$m from the primary vertex. Taking into account the preliminary nature of our study, we also consider the detector to have a spherical symmetry, instead  of a cylindrical one.

\paragraph{Decays outside the detector}

Also in this case, the precise value of the decay length of the RH neutrinos 
needed for it to be considered detector stable
depends on the specific geometry of the detector. We then consider as detector stable, RH neutrinos which decay more than $5\,$m away from the primary vertex.

\vspace{0.6cm}

\begin{figure}[t!]
\begin{center}
\includegraphics[width=0.49\textwidth]{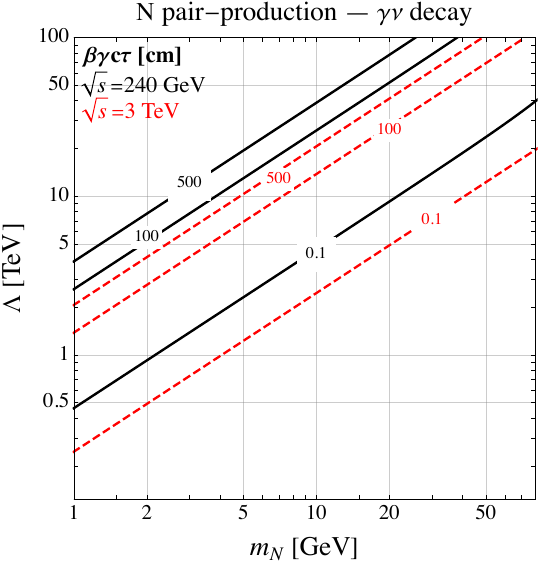}\hfill
\includegraphics[width=0.49\textwidth]{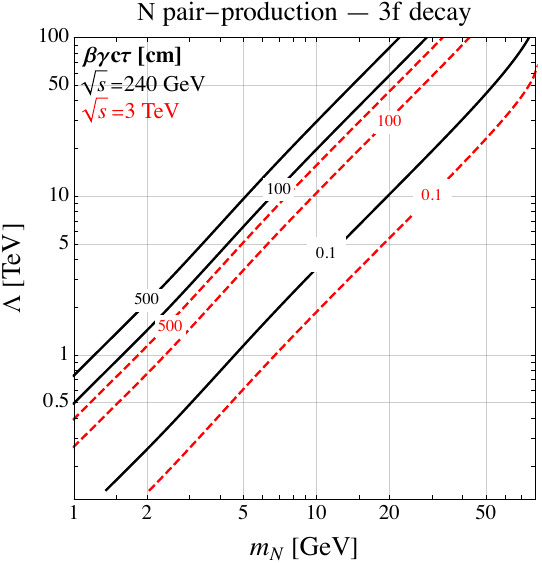}
\caption{\small Isocontour of $\beta\gamma c\tau$ in cm from exclusive $N\to \nu\gamma$ (left) and $N\to 3f$ (right) decays induced by $d>4$ operators for $\sqrt s=240\;$GeV (black) and $\sqrt s=$3\;TeV (red). Neutrino pair-production is assumed.
We show the results in the limit of negligible active-sterile mixing. }
\label{fig:ctau_ONB6}
\end{center}
\end{figure}

\noindent The decay length in the laboratory frame $\beta\gamma c\tau$ can be readily obtained for the two dominant $N$ production modes that will be discussed in Sec.~\ref{sec:prod}, {\emph{i.e.}} pair-production and single-production from four-Fermi operators. The $\beta \gamma$ factor is fixed by the kinematic of the process and reads
\be\label{bgctau}
\beta\gamma  = \frac{\sqrt s}{2 m_N}
\begin{cases}
\sqrt{1-\frac{4m_N^2}{s}}  \quad\quad {\rm Pair-production} \\
\\
\;1-\frac{m_N^2}{s}  \,~~~\quad\quad {\rm Single-production} 
\end{cases} \ .
\ee
As discussed in the previous section, in the region where the RH decay width is dominated by the $d>4$ operators, two decays compete: $N\to \nu \gamma$ and $N\to 3f$. As an example, we show in Fig.~\ref{fig:ctau_ONB6} the isocontours of $\beta\gamma c \tau$ for the case of exclusive $\nu \gamma$ (left) and $3f$ (right) decay, fixing $\sqrt s=240\;$GeV and 3\;TeV and considering the pair-production case. 
These lifetimes 
are dominated by mixing unsuppressed operators and thus do not strongly depend on the mixing angle. 
As in Sec.~\ref{sec:decay_dominate},
the dependence on $\xi$ is extremely mild. 
The case of single-production is qualitatively similar, 
with more pronounced differences appearing for large $m_N$ in the case $\sqrt s=m_Z$.
From the figures we see that the RH neutrino can have, for both final states, a prompt, displaced and stable behaviour, depending on the values of $m_N$ and $\Lambda$ considered, although a detector stable $N$ can only arise for $m_N \lesssim 20\;$GeV for $\Lambda \lesssim 100\;$TeV.
Clearly, if one considers only the decay induced by mixing suppressed operators these will in general give larger values for the proper $c\tau$ decay length, which are compatible with a displaced or stable behavior for $N$ and that can be of the same order of magnitude as the one induced by the active-sterile mixing.

\section{Production modes for RH neutrinos}\label{sec:prod}

At the renormalizable level, RH neutrinos are produced only via their mixing with the active neutrinos, while at $d=5$ two different production mechanisms arise: one from an exotic decay of the Higgs boson and one from the exotic decay of the $Z$ boson. These have been studied in 
\cite{Barducci:2020icf}, where the $N$ were considered to decay only via mixing, being this the dominant mechanism for $d\le 5$. The inclusion of $d=6$ operators brings new production modes for RH neutrinos. The main mechanisms can be divided in two categories. 
\begin{itemize}
\item [{\emph{i}})] Single- and pair-production of $N$ via four-Fermi operators,
\item [{\emph{ii}})] $N$ production via $Z,W$ and $h$ decay from $d=6$ operators involving the Higgs boson.
\end{itemize}

In this work  we focus on production via four-Fermi operators while we leave the analysis of the production from SM boson decay for future work.

\subsection{Single and pair-production of $N$ via four-Fermi operators}

At lepton colliders there are three four-Fermi operators that can produce RH neutrinos.
The ${\cal O}^6_{Ne}$ and ${\cal O}^6_{NL}$ operators 
generate the process
$\ell^+ \ell^- \to N_i  N_j$  with a rate
\be\label{eq:xs_pair}
\sigma_{{\cal O}^6_{Ne,NL}} = \frac{1}{48\pi \Lambda^4}(s-4 m_N^2)\sqrt{1-\frac{4m_N^2}{s}} 
\sim 
150\;{\rm fb} \left(\frac{\sqrt s}{240\;{\rm GeV}} \right)^2 \left(\frac{1\;{\rm TeV}}{\Lambda} \right)^4  \ ,
\ee
while the operator ${\cal O}^6_{LNLe}$ induces $\ell^+\ell^- \to \nu_\alpha N_i$ with a rate
\begin{align}\label{eq:xs_single}
\sigma_{{\cal O}^6_{LNLe}} = \frac{1}{24\pi \Lambda^4} s \left(1-\frac{m_N^2}{s}\right)^2\left(1+\frac{m_N^2}{8s} \right)
\sim 
300\;{\rm fb} \left(\frac{\sqrt s}{240\;{\rm GeV}} \right)^2 \left(\frac{1\;{\rm TeV}}{\Lambda} \right)^4 \ ,
\end{align}
where the numerical approximation is valid in the massless limit. 
In both cases we have set to unity the Wilson coefficient of the operator inducing the process and assumed fixed flavors. Appropriate multiciplicity factors must be included to compute the inclusive cross-sections in all flavors.

\begin{figure}[t!]
\begin{center}
\includegraphics[width=0.48\textwidth]{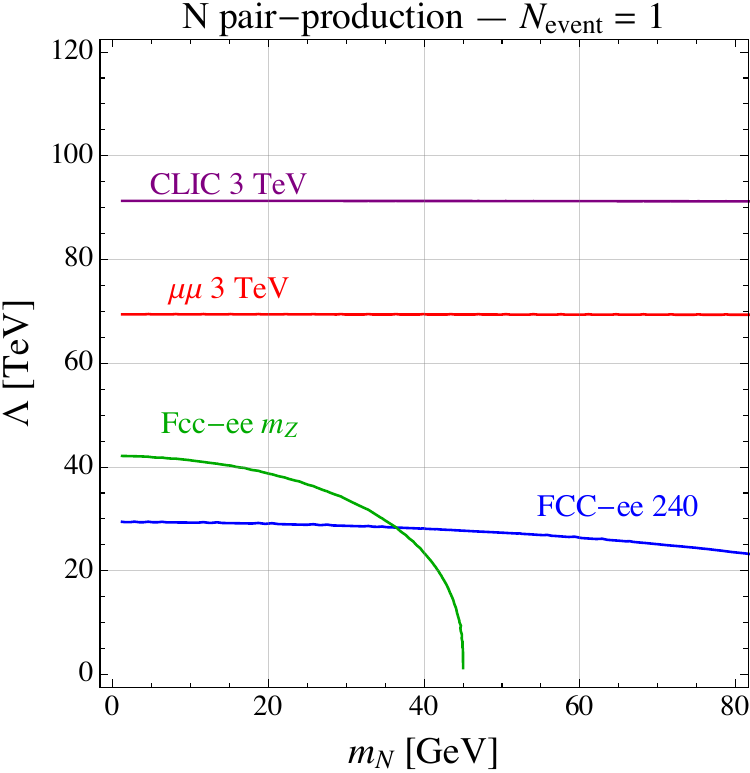}\hfill
\includegraphics[width=0.48\textwidth]{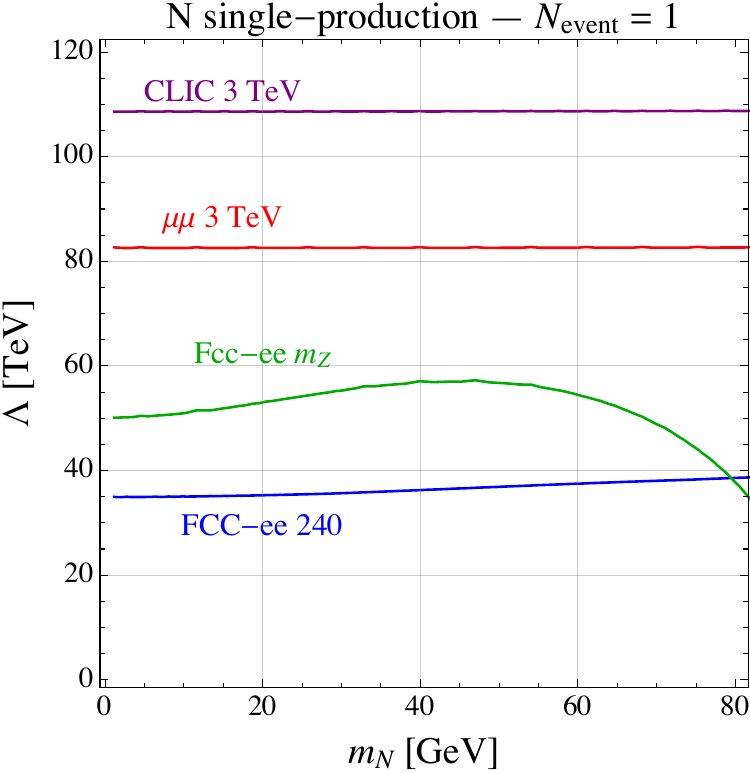}\hfill
\caption{\small Isocontours for the production of one $e e\to N N$ event from ${\cal O}^6_{Ne,NL}$ (left) and one  $e e\to \nu N$ event from ${\cal O}^6_{LNLe}$ (right) for different collider options, whose characteristics are reported in Tab.~\ref{tab:colliders}.}
\label{fig:one_event_4f}
\end{center}
\end{figure}

As a preliminary indication, we can ask what is the maximum scale $\Lambda$ that can be tested by requiring the production of at least one signal event before enforcing any BR factor and selection acceptance.  As mentioned in Sec.~\ref{sec:colliders},
we take as benchmark colliders 
the FCC-ee at $\sqrt s=240\;$GeV, 
the FCC-ee  at the $Z$ pole, a $\mu\mu$ collider with  $\sqrt s=3\;$TeV and CLIC at $\sqrt{s} = 3$ TeV. For all these options, the considered integrated luminosities are reported in Tab.~\ref{tab:colliders}. The maximum scales that can be tested are show in Fig.~\ref{fig:one_event_4f}, where the left and right panel are for $N$ pair- and single-production respectively.
By comparing this result with Fig.~\ref{fig:mix_vs_d} we see that, for light $N$  in the majority of the allowed parameter space that can be tested, the decay of the RH neutrino will proceed via higher-dimensional operators while for heavier $N$ the decay might also proceed via active-sterile mixing. 

Even if produced via a four-Fermi operator, the heavy $N$ can nevertheless decay into a $\gamma\nu$ final state. 
For instance, four-Fermi operators of the form $(\bar N \gamma^\mu N)(\bar f \gamma_\mu f)$ will induce an unsuppressed pair-production cross-section $e^+ e^-\to NN$ and a decay $N \to\nu \ell \bar\ell$ which, being mixing-suppressed, will typically be subdominant. In addition, this will also always happen for $m_N \lesssim 15$ GeV, see Eq.~\eqref{eq:na_vs_3f}.
When singly produced via ${\cal O}^6_{LNLe}$ instead the main decay mode can still be the one into a SM neutrinos and a $\gamma$ if only ${\cal O}^6_{LNB}$ and ${\cal O}^6_{LNLe}$ are present and $m_N\lesssim 100\;$GeV, while it will decay into $\ell^+\ell^-\nu$ otherwise. 

All together, it is clear that one can envisage configurations where both the $\nu\gamma$ and $3f$ decays can dominate  when the RH neutrinos are produced via four-Fermi operators. In order to be concrete, we thus analyze the two possible signatures in turn separately, assuming a 100\% exclusive decay for each mode and separately considering the possibility of a prompt, displaced and collider stable behavior.

\section{$N$ prompt decay}\label{sec:prompt}

As shown in Fig.~\ref{fig:ctau_ONB6} the RH neutrino can promptly decay into a $\nu\gamma$ and $3f$ final state in all the $N$ mass range of our interest if $\Lambda$ is sufficiently small. We start by considering the exclusive $N\to \nu\gamma$ decay, moving then to the $N\to 3f$ one for both $N$ single- and pair-production.

\subsection{Decay $N\to \nu \gamma$}

When the dominant decay mode is the one into a SM neutrino and a photon we consider the following processes for pair- and single-production of $N$
\be\label{eq:nugamma_pair}
\ell^+ \ell^-  \to  \bar N N \to \nu \nu \gamma\gamma = \slashed E_T + 2\gamma \ , 
\ee
and
\be\label{eq:nugamma_single}
\ell^+ \ell^-  \to  \bar N \nu \to \nu\nu \gamma = \slashed E_T + \gamma  \ .
\ee

In the case of $N$ pair-production, Eq.~\eqref{eq:nugamma_pair}, the final state consists in a pair of $\gamma$ and $\slashed E_T$. Two operators can mediate the $N$ pair-production: ${\cal O}^6_{Ne,NL}$ whose cross section is reported, for each process, in Eq.~\eqref{eq:xs_pair}. For simplicity, and being conservative, we assume that only one of the two operators is present and only one pair of RH neutrino is produced. 
When the RH neutrino is singly produced, Eq.~\eqref{eq:nugamma_single}, the final state consist of a single photon and $\slashed E_T$. Only one operator can mediated this process, ${\cal O}^6_{NeNL}$ , whose cross-section is reported in Eq.~\eqref{eq:xs_single}.  We have implemented\footnote{{\tt MadGraph5\_aMC\@NLO} is not compatible with four-Fermi operators involving Majorana particles. We have thus implemented an appropriate $\nu$SMEFT UV completion  and fixed the masses of the relevant state to a value such that the EFT description applies.
} the relevant higher-dimensional operators in the {\tt Feynrules} package~\cite{Alloul:2013bka} and exported it under the {\tt UFO} format~\cite{Degrande:2011ua}
in order to  generate parton level signal events with {\tt MadGraph5\_aMC\@NLO}~\cite{Alwall:2014hca}. Events has been then analysed with the {\tt MadAnalysis5} package~\cite{Conte:2012fm,Conte:2014zja,Dumont:2014tja}. The irreducible SM backgrounds $\ell^+ \ell^- \to \gamma\gamma \slashed E_T$ and $\ell^+ \ell^- \to \gamma \slashed E_T$ have been generated with the same prescription. 
At the analysis level, we require the photon to be reconstructed with 
$|\eta^\gamma|<2.44$  and, for the pair-production case, that they are separated as $\Delta R(\gamma\gamma)>0.1$. We enforce the following cut on the photon(s): in the pair-production case we apply $p_T^\gamma>80\;$GeV, 20\;GeV and 300\;GeV for the FCC-ee at $\sqrt s=240\;$GeV, the FCC-ee at $\sqrt s=m_Z$ and CLIC and the $\mu\mu$ collider at 3\;TeV respectively. In the single-production case we apply instead $p_T^\gamma>20\;$GeV, 20\;GeV and 300\;GeV for the same three collider options. The statistical significance is evaluated in units of standard deviations as $S/\sqrt{S+B}$ where $S$ and $B$ are the final number of signal and background events respectively.  

\begin{figure}[t!]
\begin{center}
\includegraphics[width=0.48\textwidth]{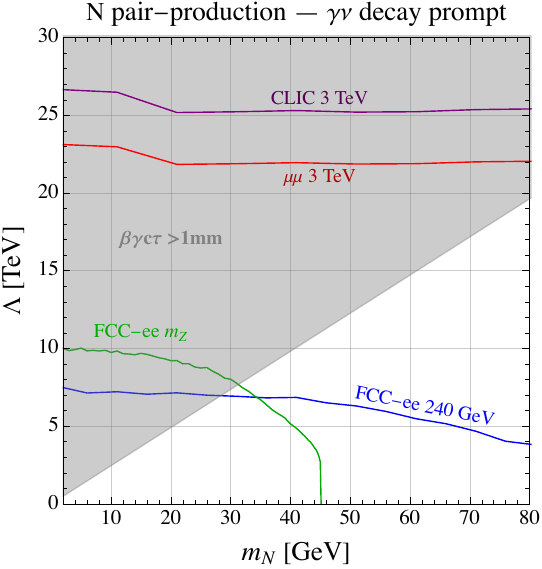}
\hfill
\includegraphics[width=0.48\textwidth]{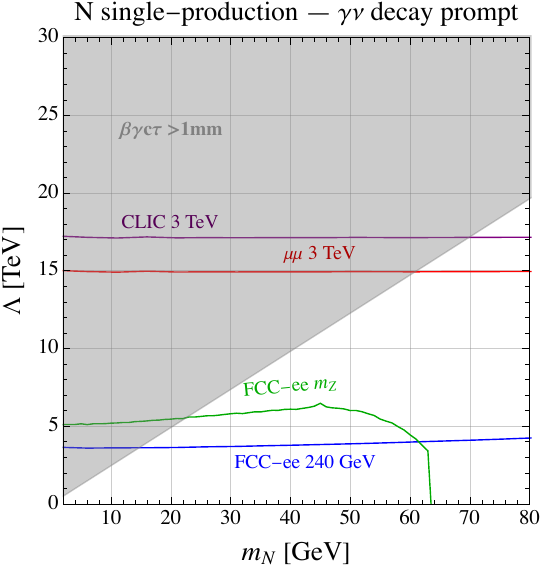}
\caption{\small 
95\% CL exclusion limit for the prompt decay into $\nu\gamma$ for 
$N$ pair-production (left) and single-production (right) for various collider options. Also indicated is the region where the decay cannot be prompt so that the described analysis doesn't apply. See text for more details.
}
\label{fig:nugamma_pair_prompt}
\end{center}
\end{figure}
We then show in Fig.~\ref{fig:nugamma_pair_prompt} the  the 95\% confidence level (CL) exclusion contours for the four collider options for the pair-production (left) and single-production (right) cases respectively. In the figures the gray shaded area is the region with $\beta\gamma c\tau>0.1\;$cm, that is where the RH neutrinos do not decay promptly and the analysis doesn't apply.  This region is conservatively shown for $\sqrt s=3\;$TeV and is smaller for lower collider energies, see Fig.~\ref{fig:ctau_ONB6}.
In the pair-production case we observe that the FCC-ee running at the $Z$ mass has a higher sensitivity to this scenario with respect to the FCC-ee running at $\sqrt s=240\;$GeV, thanks to the higher integrated luminosity of the first option. In the region where the prompt analysis applies, the bound reaches its maximum at around $m_N\sim 30\;$GeV, then depleting at the mass threshold for $N$ pair-production, where the 240\;GeV run of FCC-ee will retain a sensitivity up to $\Lambda \sim 5\;$TeV. Note that the bound on $\Lambda$ from Higgs precision measurements, see Eq.~\eqref{eq:precision_bound}, partially covers these regions if  the ${\cal O}^6_{LNH}$ operator is switched on. 
On the other side a $\mu\mu$ collider running at $\sqrt s=3\;$TeV will be able to test in principle up to $\Lambda\sim 20\;$TeV, while CLIC at the same center of mass energy will be able to test scales up to $\Lambda \sim 25$ TeV. However only lower scales will be effectively tested by this analysis since for higher values of $\Lambda$ the RH neutrinos will not decay promptly. 
We also note that the reach is dramatically reduced with respect to the maximal one, left panel of Fig.~\ref{fig:one_event_4f}, due to the non negligible SM background for this process. On this respect the limits obtained in Fig.~\ref{fig:nugamma_pair_prompt} can however be  considered as conservative and can be improved by dedicated background treatment and reduction, thus increasing the overall reach on $\Lambda$ in a realistic analysis. The results in the single-production case are qualitatively similar, albeit slightly weaker, with respect to the pair-production scenario, due to the higher rate for the SM background.

\subsection{Decay $N\to 3 f$}

\begin{figure}[t!]
\begin{center}
\includegraphics[width=0.48\textwidth]{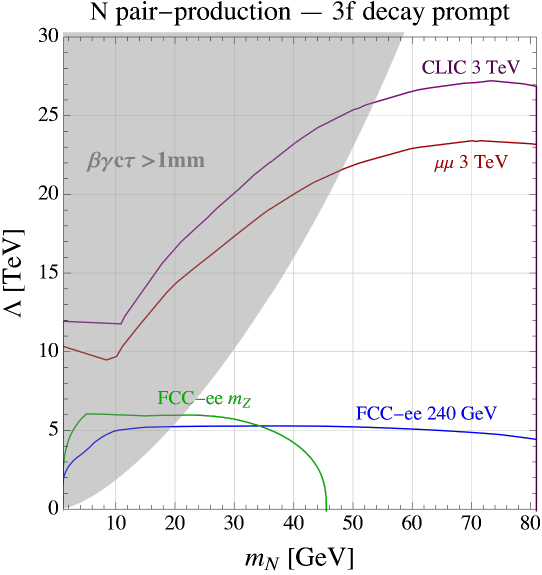}\hfill
\includegraphics[width=0.48\textwidth]{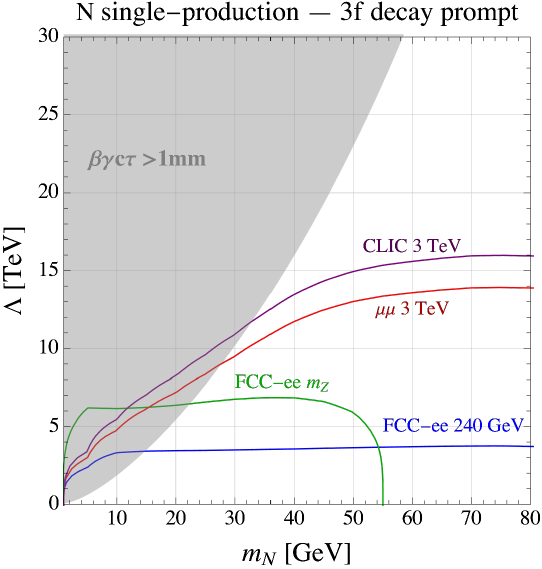}
\caption{\small 
95\% CL exclusion limit for the prompt decay into $3f$ for 
$N$ pair-production for various collider options. Also indicated is the region where the decay cannot be prompt so that the described analysis doesn't apply. See text for more details.
}
\label{fig:3f_pair_prompt}
\end{center}
\end{figure}

When the dominant decay mode is the one into three SM fermions, we consider the following processes for pair- and single-production of $N$:
\be
\ell^+ \ell^-  \to  N N \to 6f 
\ee
and
\be
\ell^+ \ell^-  \to  N \nu \to  3f \slashed E_T ,
\ee
where the fermion final state could also include quarks. These final state are similar to the one that arises by singly or pair-produced $N$ that decay via mixing, albeit with a different kinematics\footnote{
In practice, we consider a scenario where the decay is triggered by the ${\cal O}^6_{NeH}$ operator, which mediate $N\to \ell W^*$. Not being mixing nor loop suppressed, this decay is the dominant one even when the ${\cal O}^6_{LNLe}$ operator that mediate single-production and that can trigger $N\to \ell\ell \nu$ is switched on.
}. For the pair-production case we focus on the following process with a pair of same-sign (SS) leptons, which is expected to be particularly clean 
\be
e^+ e^- \to N N \to \ell^{\pm} \ell^{\pm} 4q \ ,
\ee
where the four quarks arise from the virtual $W$ decay and can be in any flavor combination. As for the SM background, we follow the same procedure of~\cite{Barducci:2020icf} and compute the  SM background $\ell^+ \ell^- \to \ell^+ \ell^- 4 q$, correcting it for a (flat) lepton charge mis-identification probability factor of $\epsilon_{{\rm misID}}^\ell = 10^{-3}$~\cite{ATLAS:2019jvq}, {\emph{e.g.}} we compute the background yield as
$\sigma_{\ell^+\ell^- 4q}\times 2 \times \epsilon_{{\rm misID}}^\ell (1-\epsilon_{{\rm misID}}^\ell)$. 
 At the analysis level, we require $p_T^\ell>2.5\;$GeV, $p_T^j>5\;$GeV, $|\eta^\ell|<2.44$, $|\eta^j|<2.4$ and $\Delta R>0.1$\footnote{We require $\Delta R>0.05$ in the case of CLIC and the 3 TeV $\mu\mu$ collider.} between the two leptons and a lepton and jet pair\footnote{We do not require jets to be isolated between themselves as we assume that they can be reconstructed as fat-jet objects.}. We furthermore consider the correct mass dependent SS branching ratio from the $N$ decay induced by  ${\cal O}^6_{NeH}$. 
 We thus obtain the 95\% CL exclusion limit shown in the left panel of Fig.~\ref{fig:3f_pair_prompt}, where we see that the FCC-ee will be able to test roughly $\Lambda \sim 5\;$TeV in the whole considered $N$ mass range for both runs at the $Z$ pole mass and at $\sqrt s=240\;$GeV, while the high-energy colliders will be able to test up to $\Lambda\sim 20-25\;$TeV, although only in a smaller region the RH neutrino will decay promptly. For the single-production case, whose results are shown in the right panel of Fig.~\ref{fig:3f_pair_prompt}, we study the single lepton channel
\be
\ell^+ \ell^- \to N \nu \to \ell 2q \slashed E_T \ ,
\ee
and the corresponding SM irreducible background. Other than the same basic selection cuts imposed in the pair-production case, we further impose a requirement on the missing transverse energy $\slashed E_T^{\rm miss} >  \sqrt{s}/3$. This is motivated by the fact 
that in the signal case the light active neutrino carries away $\sim 50\%$ of the available center of mass energy, while this is not the case for the background processes, for which the $\slashed E_T$ distribution peaks at lower values.

\section{$N$ displaced decay}\label{sec:displaced}

 \begin{figure}[t!]
\begin{center}
\includegraphics[width=0.48\textwidth]{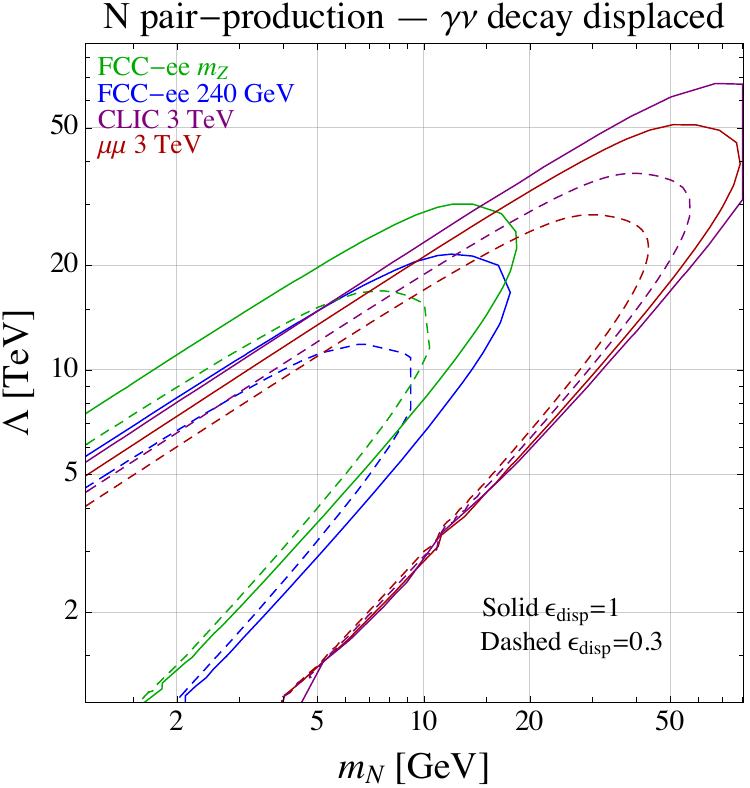}\hfill
\includegraphics[width=0.48\textwidth]{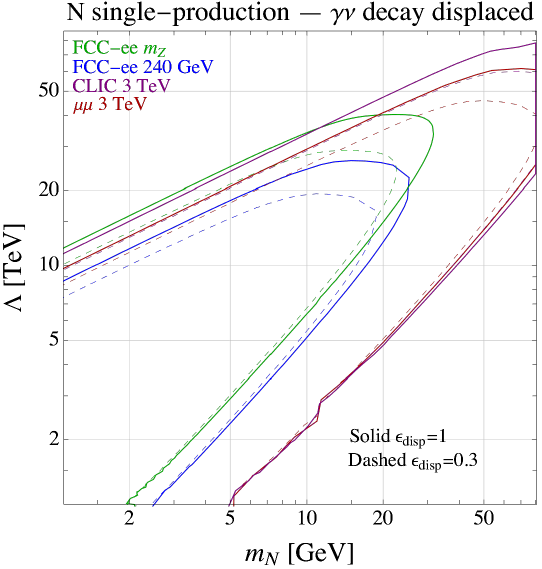}
\caption{\small 
95\% CL exclusion limits for the displaced decay into a $\nu\gamma$ final state for pair produced (left) and singly produced (right) RH neutrinos. The solid lines correspond to $\epsilon_{\rm disp}=1$ while the dashed ones to $\epsilon_{\rm disp}=0.3$ 
}
\label{fig:displaced_na}
\end{center}
\end{figure}

\begin{figure}[t!]
\begin{center}
\includegraphics[width=0.48\textwidth]{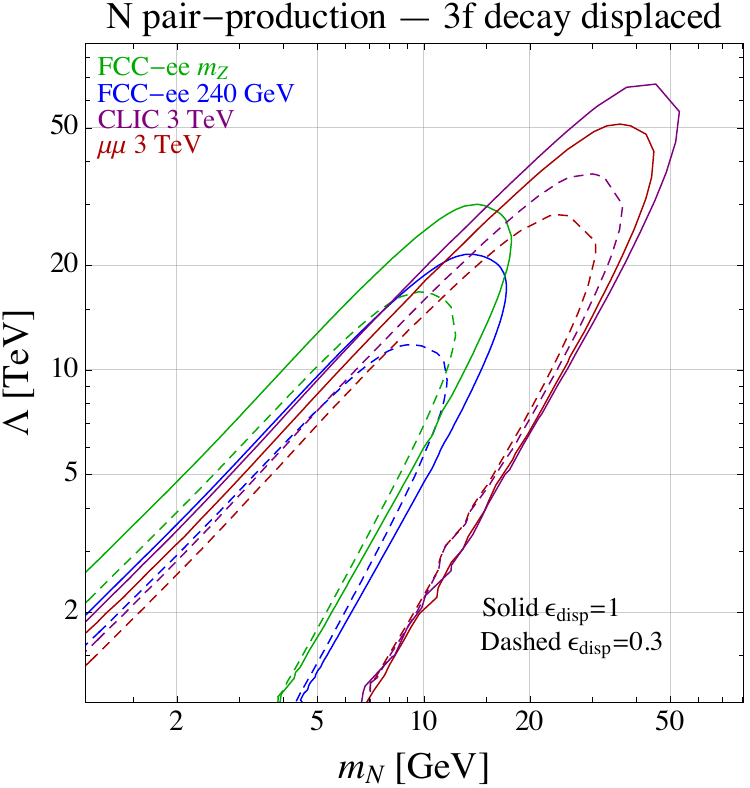}\hfill
\includegraphics[width=0.48\textwidth]{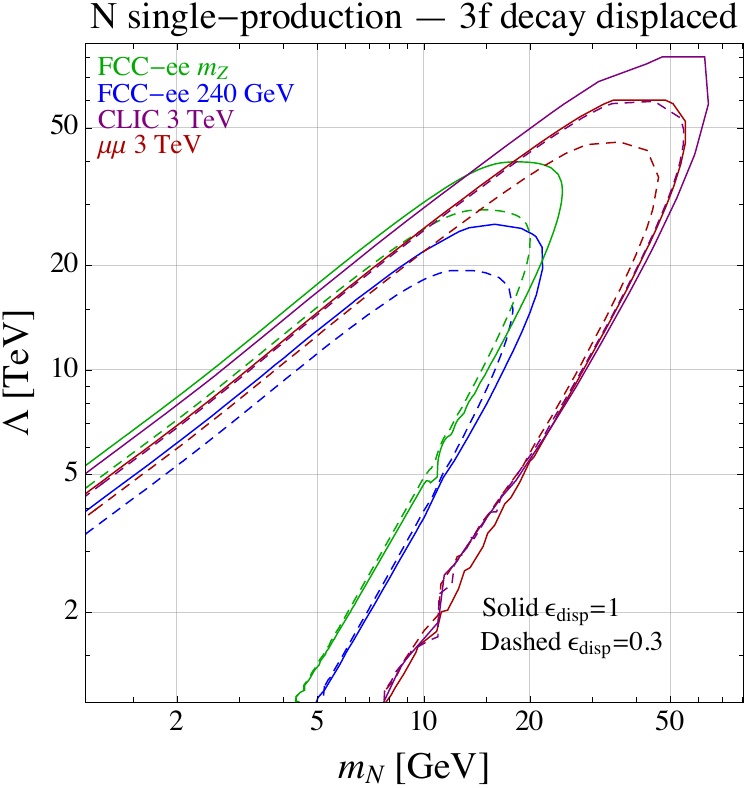}
\caption{\small 
95\% CL exclusion limits for the displaced decay into a  $3f$ final state  for pair produced (left) and singly produced (right) RH neutrinos. The solid lines correspond to $\epsilon_{\rm disp}=1$ while the dashed ones to $\epsilon_{\rm disp}=0.3$. 
}
\label{fig:displaced_3f}
\end{center}
\end{figure}

We now study the sensitivity for RH neutrinos decaying with a displacement which, as discussed in Sec.~\ref{sec:lifetime}, we take to be between 1\;cm and 100\;cm from the primary vertex. The final event yield for having reconstructed displaced events is parametrized as
\be\label{eq:displaced}
N_s = \sigma_{\rm prod} \times \epsilon^n_{P_{\Delta L}}\times \epsilon^n_{{\rm disp}}\times {\cal L} \ ,
\ee
where $\sigma_{\rm prod}$ is the pair-production or single-production cross section for $N$ and ${\cal L}$ denotes the total integrated luminosity.
 $\epsilon_{P_{\Delta L}}$ represents the acceptance for having a RH neutrino decaying within a certain displacement from the primary vertex. This can be computed from the exponential decay law, taking into account the Lorentz time dilation factor.
We then assign a probability for having the RH neutrino decaying at a distance $\Delta x=x_f-x_i$ which reads
\be
{\cal P}(x_i,x_f)=e^{-\frac{x_i}{\beta\gamma c \tau}}-e^{-\frac{x_f}{\beta\gamma c \tau}} \ ,
\ee
where the $\beta\gamma c\tau$ factors are reported in Eq.~\eqref{bgctau}  for pair-production and single-production cases, for which the parameter $n$ in Eq.~\eqref{eq:displaced} takes the value of 2 and 1 respectively. This means that for the pair-production case we ask to reconstruct both RH neutrinos as decaying displaced.
With $\epsilon_{{\rm disp}}$ we instead parametrize the acceptance for reconstructing the displaced decaying neutrino, which depend on the actual detector design and performances, and which therefore we assume as a free extra parameter in the analysis. 
The irreducible SM background is expected to be negligible on the considered decay lengths and we thus work in the zero background hypothesis. We then show the expected 95\% CL exclusion limits, now obtained by requiring $N_s>3$, in Fig.~\ref{fig:displaced_na} and Fig.~\ref{fig:displaced_3f} for the pair-production and single-production cased under the assumption of exclusive $\nu\gamma$ and $3f$ decay respectively. The solid and dashed lines correspond to the choice $\epsilon_{\rm disp}=1$ and 0.3 respectively, while the different colors represents the different collider options.

From the results we observe that a displaced analysis at the FCC-ee running at $\sqrt s=240\;$GeV
can be sensitive to ${\cal O}(10\;{\rm TeV})$ NP scale with a 30\% efficiency on the reconstruction of the displaced for $m_N\lesssim 10\;$GeV in the pair-production case, while a higher reach can be attained in the single-production scenario. The FCC-ee 
 running at the $Z$ pole mass can slightly increase these reach due to the large integrated luminosity, while the 3\;TeV collider prototypes can reach up to $\Lambda \sim 50-60\;$TeV for $m_N\sim 40\;$GeV.

\section{Detector stable $N$}\label{sec:stable}

Finally, we discuss the possibility of detector stable RH neutrinos, 
{\it e.g.} the case in which the decay happens 
more than 500\;cm away from the interaction vertex. In this case, both pair-production and single-production give rise to a totally invisible final state. This process can be targeted through the emission of an initial state photon, producing a mono-$\gamma$ signature, $\ell^+\ell^- \to \gamma \slashed E_T$, which has as SM background $\ell^+ \ell^-\to \nu\bar \nu \slashed E_T$.
 In~\cite{Habermehl:2020njb} exclusion prospects for various four-Fermi operators producing a weakly interacting massive particle  dark matter candidate were given using a full detector simulation of the International Linear Detector prototype for the International Linear Collider. Moreover, rescaling factors for different collider energies, luminosities and beam polarizations where provided.
Based on these results at the FCC-ee with 5\;ab$^{-1}$ of integrated luminosities, cutoff scales up to $\Lambda \sim 1.5\;$TeV can be tested in the pair-production case. In the single-production case the cross-section is larger than in the pair-production case but the photon spectrum is expected to be more similar to the SM due to the presence of only one heavy particle in the final state. Overall we thus expect the exclusion reach on $\Lambda$ to be similar to the one of the pair-production case. However for such low scale the RH neutrino $N\to \gamma\nu$ decay happens inside the detector, see Fig.~\ref{fig:ctau_ONB6}, unless there is a cancellation among the $\alpha_{LNB}$ and $\alpha_{LNW}$ Wilson coefficient, see Eq.~\eqref{eq:nugamma}. If the dominant decay is $N\to 3f$ instead, the RH neutrino can be stable on detector lengths if $\Lambda > 750\;$GeV and $m_N<2\;$GeV, so that the
derived limit of $1.5\;$TeV applies. 
 
For higher center of mass energies we can again use as a guidance the results of ~\cite{Habermehl:2020njb}. Here the derived reach of CLIC at $\sqrt s=3\;$TeV with 1\;ab$^{-1}$ of integrated luminosity is $\Lambda \sim 10\;$TeV. For CLIC and the 3\;TeV $\mu\mu$ collider at the same center of mass energy we expect a reach in the same ballpark, although a dedicated study is required for a quantitative assessment. By again a comparison with Fig.~\ref{fig:ctau_ONB6} we see that a reach of $10\;$TeV on $\Lambda$ will be able to prove detector stable RH neutrinos up to $5\;$GeV and 10\;GeV if the only available decay mode is the one into $\nu\gamma$ and $3f$ respectively.

\section{Conclusions}\label{sec:conclusions}

\begin{figure}
\begin{center}
\includegraphics[width=0.48\textwidth]{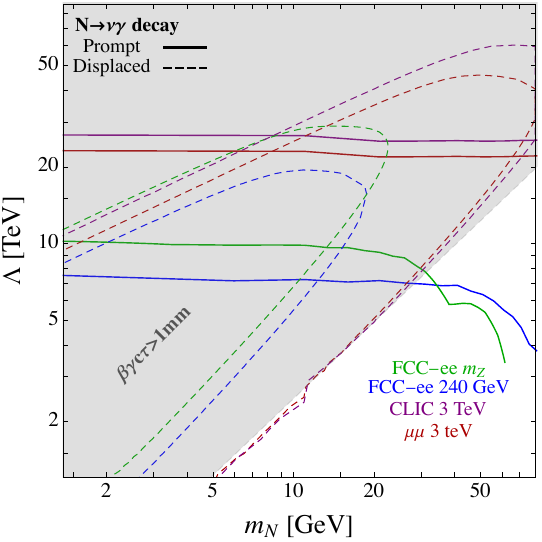}
\hfill
\includegraphics[width=0.48\textwidth]{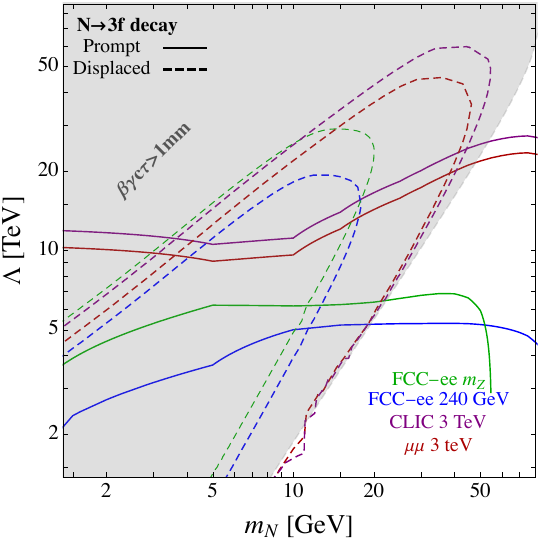}
\caption{Summary of the main results of this paper. Left panel: 95\% C.L. exclusion limits for RH neutrinos decaying via $N \to \nu \gamma$. Right panel: same as in the left panel, but for RH neutrinos decaying via $N \to 3f$. In the gray region the decay is displaced. The solid lines show the sensitivity for prompt decays (valid in the white region), while the dashed lines show the sensitivity for displaced decays. The latter curves are drawn considering an efficiency of reconstruction of $30\%$. }
\label{fig:moneyplot}
\end{center}
\end{figure}

In this paper we have considered the $\nu$SMEFT and studied how the RH neutrinos $N$ production and decays may be affected by the inclusion of $d=6$ operators. More specifically, we have studied the reach of future Higgs factories machines on the cutoff scale $\Lambda$ at which the EFT is generated. We focused on four representative machines: the FCC-ee at two different center-of-mass energies, $\sqrt{s} = 90$ GeV and $\sqrt{s} = 240$ GeV, CLIC at a center of mass energy of 3\;TeV and a representative muon collider with $\sqrt{s} = 3$ TeV. 

The complete list of non-redundant $d=6$ operators is presented in Tab.~\ref{tab:D6_operators}. At the level of production, the $d=6$ operators induce either $N$ pair- or single-production. On the other hand, at the level of decays, they induce the modes $N \to \nu \gamma$ and $N \to 3f$, where various fermions combinations are possible. The former will dominate for RH neutrino masses $m_N \lesssim 15$ GeV, while the latter will dominate for larger masses, unless the only operators switched on induce a mixing-suppressed decay. Even more interestingly, depending on the RH neutrino mass and on the cutoff scale $\Lambda$ at which the EFT is generated, the decays can be prompt, displaced or the RN neutrinos can be collider stable. The phenomenology crucially depends on their decay behavior and we have analyzed in detail all three possibilities. Our analysis is reported in Sec.~\ref{sec:prompt}, Sec.~\ref{sec:displaced} and Sec.~\ref{sec:stable} for the three possible RH neutrinos lifetime. We then summarize the results for convenience in Fig.~\ref{fig:moneyplot}, in which, for the Higgs factories considered in this work, we show the 95\% C.L. exclusion on the scale $\Lambda$ as a function of $m_N$. We consider RH neutrino masses up to $80$ GeV. For larger masses, the $W$ boson can be produced on-shell in the $N$ decays and our analysis should be slightly modified. We postpone the analysis of such case to future work, although we do not expect major changes with respect to the results shown here. In the left panel we consider the decay channel $N \to \nu \gamma$, while in the right panel we show the results for $N \to 3 f$. In both panels, the gray region denotes the parameter space in which the RH neutrino decay is displaced. The solid lines show the exclusion (combining pair and single-production) computed with prompt decays, an analysis valid in the white region. The dashed lines, on the contrary, show the exclusion limit considering displaced decays with an efficiency of reconstruction of $30\%$. In the region of validity of the prompt analysis, the FCC-ee will be able to probe scales up to $\Lambda \sim 7$ TeV, while larger values, up to $\Lambda \sim 20-30$ TeV, can be probed with a displaced analysis. These conclusions are valid for both decay channels. In the case of the colliders at $3\;$TeV, on the other hand, scales up to $\Lambda \sim 20\div 30\;$TeV can be probed while the displaced analysis, on the other hand, allows to probe scales up to $\Lambda \sim 60$ TeV.

\section*{Acknowledgements}

We thank Pilar Hernandez and Barbara Mele for contributing at the early stage of this work and for useful discussions. E.B. acknowledges financial support from “Funda\c{c}\~ao de Amparo \`a Pesquisa do Estado de
S\~ao Paulo” (FAPESP) under contract 2019/04837-9

%%%%%%%%%%%%%%%%%
%%%	APPENDIX	   %%%		
%%%%%%%%%%%%%%%%%

 \appendix

\section{Spin averaged matrix elements for $N$ decay}\label{sec:amplitudes}

We list here the spin-averaged matrix elements $|{\overline {\cal M}}|^2 = \frac{1}{2}\sum_{\rm spins} |{\cal M}|^2$ for the three body decays of the RH neutrino via the $d=6$ operators that proceed through an off-shell boson considered in the text. The kinematics is fixed as $1\to 2,3, 4$ and we define $m_{ij}^2=(p_i + p_j)^2$. The final state SM neutrino is always considered to be massless while, depending on the simplicity of the expressions, some of the amplitudes 
are reported in the limit of vanishing masses for the other final state fermions.
From these amplitudes squared the partial widths are readily obtained as~\cite{ParticleDataGroup:2020ssz}
\be
{\rm d}\Gamma = \frac{1}{(2\pi)^3}\frac{1}{32 m_N^3} |{\overline {\cal M}}|^2{\rm d}m_{23}^2 {\rm d}m_{34}^2 .
\ee

\subsection*{Decay from ${\cal O}^6_{LNH}$: $N \to \nu H^*, H^* \to f \bar f$}
This amplitude include the contribution from both the decay modes due to the Majorana nature of the involved particles. With $m_3=m_4=m_f$ it reads 
\be
|{\overline {\cal M}}|^2 = 
\left|(\alpha_{LNH})_{\alpha i} \right|^2 \frac{9}{4} \left(\frac{v \, m_f}{\Lambda^2}\right)^2\frac{(m_N^2 - m_{34}^2)(m_{34}^2-4 m_f^2)}{(m_{34}^2-m_H^2)^2} \ .
\ee

\subsection*{Decay from ${\cal O}^6_{NH}$: $N \to \nu Z^*, Z^* \to f \bar f$}
Also in this case the amplitude includes both decay modes due to the Majorana nature of the involved particles. In the limit $m_3=m_4=m_f=0$ it reads
\begin{align}
|{\overline {\cal M}}|^2 = & 
\left| (\alpha^6_{NH})_{ij} \theta_{j \alpha} \right|^2
 \frac{1}{2}
\left(\frac{e}{s_\omega c_\omega}\right)^4(g_L^2 + g_R^2) \left(\frac{v}{\Lambda}\right)^4  \frac{m_N^2(2 m_{23}^2+m_{34}^2)-2 m_{23}^4-2 m_{23}^2 m_{34}^2 - m_{34}^4}{(m_{34}^2-m_Z^2)^2} ,
\end{align}
where
\be
g_L = t_3  - q s_w^2,  \qquad g_R = - q s_w^2 .
\ee

\subsection*{Decay from ${\cal O}^6_{NeH}$: $N \to \ell W^*, W^* \to f_3 \bar f_4$}

Here the reported amplitude is for just one of the two charged-conjugated decay modes. Setting $m_2=m_\ell=0$ and $m_{f_{3,4}}=m_{3,4}=0$
one obtains
\be
|{\overline {\cal M}}|^2 = \left| (\alpha^6_{NeH})_{i\alpha} \right|^2 \frac{g^4}{2}\left(\frac{v}{\Lambda} \right)^4 \frac{(m_N^2-m_{23}^2-m_{34}^2)(m_{23}^2+m_{34}^2)}{(m_{34}^2-m_W^2)^2} \ .
\ee

\subsection*{Decay from ${\cal O}^6_{LNB,W}$: $N \to \nu Z^*, Z^* \to f\bar f$}

Again the amplitude include the contribution from both the decay modes due to the Majorana nature of the involved particles. By setting $m_3=m_4=m_f=0$ one obtains \\
\begin{align}
|{\overline {\cal M}}|^2 = & 
4\frac{|(\alpha_{LNB}^6)_{\alpha i} c_\omega + (\alpha_{LNW}^6)_{\alpha i} s_\omega|^2}{(16\pi^2)^2} \left( \frac{e}{s_\omega c_\omega}\right)^2 \frac{v^2}{\Lambda^4}(g_L^2+g_R^2)m_{34}^2 \times \nn \\
& \qquad \times \frac{
2 m_{23}^2 (m_{23}^2+m_{34}^2) + m_N^4- m_N^2 (2 m_{23}^2+m_{34}^2)
}{(m_{34}^2-m_Z^2)^2} .
\end{align}

\subsection*{Decay from ${\cal O}^6_{LNW}$: $N \to \ell W^*, W^* \to f_3\bar f_4$}

Here the amplitude squared is for just one of the two charged-conjugated decay modes.  With all massless final states one obtains
\be
|{\overline {\cal M}}|^2 = \frac{| (\alpha^6_{LNW})_{\alpha i}|^2}{(16\pi^2)^2}  \frac{4 g^2 v^2}{\Lambda^4} \frac{m_{23}^2 m_{34}^2 (m_{23}^2+m_{34}^2)}{(m_{34}^2-m_W^2)^2} .
\ee

\subsection*{Decay from four-Fermi operators}

For four-Fermi operators of vector type involving two RH neutrinos we have that, in the limiting of vanishing final state masses,
\be
|{\overline {\cal M}}|^2 = \frac{\left| (\alpha^6)_{ij} \theta_{j \alpha} \right|^2}{\Lambda^4} \left[  (2 m_{23}^2 + m_{34}^2) m_N^2 -2 m_{23}^4 - 2 m_{23}^2 m_{34}^2 - m_{34}^4 \right] ,
\ee
where $\alpha^6$ is the appropriate Wilson coefficients. Once more, the amplitude refers to only one channel. 
The same averaged squared amplitude is obtained for the operator ${\cal O}^6_{Nedu}$ (without the mixing suppression). As for the scalar four-Fermi operators, under the same conditions as above we obtain
\be
|{\overline {\cal M}}|^2 = \frac{\left| (\alpha^6)_{i \alpha}\right|^2}{ 2 \Lambda^4} m_{34}^2 ( m_N^2 - m_{34}^2)\,  .
\ee

%%%%%%%%%%%%%%%%%
%%%	REFERENCES	   %%%		
%%%%%%%%%%%%%%%%%

\bibliographystyle{JHEP}
{\footnotesize
\bibliography{biblio}}
\end{document}